\documentclass[]{aa}  

\usepackage{diagbox}
\usepackage{amsmath}
\usepackage{graphicx}
\usepackage{natbib}
\usepackage{hyperref}
\usepackage{txfonts}

\DeclareRobustCommand{\ION}[2]{%
\relax\ifmmode
\ifx\testbx\f@series
{\mathbf{#1\,\mathsc{#2}}}\else
{\mathrm{#1\,\mathsc{#2}}}\fi
\else\textup{#1\,{\mdseries\textsc{#2}}}%
\fi}

\newcommand{\nii}{[\ION{N}{ii}]}
\newcommand{\oi}{[\ION{O}{i}]}
\newcommand{\oii}{[\ION{O}{ii}]}
\newcommand{\oiii}{[\ION{O}{iii}]}
\newcommand{\sii}{[\ION{S}{ii}]}

\newcommand{\EWHa}{EW($\rm{H}\alpha$)}
\newcommand{\WHa}{EW($\rm{H}\alpha$)}
\newcommand{\Ha}{$\rm{H}\alpha$}
\newcommand{\Hb}{$\rm{H}\beta$}

\newcommand{\HII}{\ion{H}{ii}}

\begin{document} 

\title{Beyond diagnostic-diagrams: A critical exploration on the classification of ionization processes}

\author{S.~F.~S\'anchez\inst{\ref{uname},\ref{iac}}
   \and C.~Mu\~noz-Tu\~n\'on\inst{\ref{iac},\ref{ll}}
   \and J. S\'anchez Almeida\inst{\ref{iac},\ref{ll}}
   \and O.~Gonz\'alez-Mart\'\i n\inst{\ref{irya},\ref{iaa}}
   \and E.~P\'erez\inst{\ref{iaa}}   
 }

 \institute{Instituto de Astronom\'ia, Universidad Nacional Auton\'oma de M\'exico, A.P. 106, Ensenada 22800, BC, Mexico \label{uname}
 \and  Instituto de Astrof\'\i sica de Canarias, La Laguna, Tenerife, E-38200, Spain \label{iac}
 \and Departamento de Astrof\'\i sica, Universidad de La Laguna, Spain \label{ll}
 \and Instituto de Radioastronom\'\i a and Astrof\'\i sica (IRyA-UNAM), 3-72 (Xangari), 8701, Morelia, Mexico \label{irya}
 \and Instituto de Astrof\'\i sica de Andaluc\'\i a (IAA-CSIC), Granada, E-18008, Spain \label{iaa}
 }


   \date{Received ---, 2025; accepted ---, 2025}


  \abstract
  { Diagnostic diagrams based on optical emission
  lines, especially the classical BPT diagrams, have long been used to
  distinguish the dominant ionisation mechanisms in galaxies. However,
  these methods suffer from degeneracies and limitations, in
  particular when applied to complex systems such as galaxies, where multiple ionisation sources coexist.}
   {We aim to critically assess the effectiveness of
  commonly used diagnostic diagrams in identifying star-forming
  galaxies, retired galaxies (RGs), and  active galactic nuclei (AGNs).
  We also explore alternative diagnostics and propose a revised
  classification scheme to reduce misclassifications and better
  reflect the physical mechanisms ionizing gas in galaxies.}
   {Using a comprehensive sample of nearby galaxies
  from the NASA-Sloan Atlas (NSA) cross-matched with Sloan Digital Sky Survey (SDSS) spectroscopic data,
  we define archetypal subsamples of late-type/star-forming galaxies,
  early-type/retired galaxies, and multi-wavelength selected AGNs. We
  evaluate their distribution across classical and more recent diagnostic
  diagrams, including the WHaN, WHaD, and a newly proposed WHaO
  diagram, which combine \Ha\ equivalent width with additional
  indicators (\nii/\Ha, $\sigma_{\rm H\alpha}$ and \oiii/\oii, respectively). We quantitatively compare the resulting classification across
  multiple schemes.}
   {Classical BPT diagrams systematically overestimate
  the number of star-forming galaxies ($\sim$10\%) and misclassify a significant
  fraction of AGNs (up to 45\%) and RGs (up to 100\%). Diagrams incorporating the equivalent width of \Ha, such
  as WHaN, WHaD, or WHaO, yield more reliable separations ($\sim$20\% of AGNs and $\sim$15\% of RGs erroneously classified). A new classification scheme
  based on \EWHa\ thresholds and concordant WHaD/WHaO results
  achieves improved purity for all classes ($\sim$8-25\% sources erroneously classified) and better alignment with
  known physical properties.}
   {The widely used BPT-based classifications fail to
  accurately distinguish between ionisation mechanisms, especially in
  galaxies hosting low-luminosity AGNs or retired stellar
  populations. Updated schemes incorporating \EWHa\ and
  complementary diagnostics, with their own limitations, provide a more accurate view of galaxy
  ionisation and should be adopted in future studies of galaxy
  populations and evolution.}

   \maketitle
%
\section{Introduction}
\label{sec:intro}

The interstellar medium (ISM) in galaxies can be ionized by very
different mechanisms which are associated with a variety of physical
processes. According to \citet{ARAA}, the most relevant ones are 
photo-ionization by  (i) OB young and massive stars
in recent star-formation (SF) events
\citep[e.g.][]{strom39,osterbrock92}, that comprises the classical
\HII\ regions, (ii) hot evolved low-mass stars/post-Asyntotic Giant Branch  stars
\citep[HOLMES/p-AGB, e.g.,][]{binette94,flor11}, observable in
non-starforming/retired galaxies (RGs) and regions within them
\citep[][]{sign13,belfiore17a} , (iii) active galactic nuclei (AGNs)
produced by the gas accretion into central super-massive black holes in
certain galaxies \citep[][]{sand65,urry95},(iv) ionization produced by
shocks at local or global scales, in particular those in
 high-velocity galactic outflows from  strong nuclear
star-formation processes and/or
AGNs\ \citep[e.g.][]{veilleux2005,carlos19}, (v) optical jets in
 AGNs\ \citep[e.g.][]{carlos17}, (vi) low velocity
outflows/inflows \citep[e.g.][]{dopita96,kehrig,roy18} and (vii)
supernovae remnants \citep[e.g.][]{cid21}. Low velocity shocks  and
HOLMES/p-AGB ionization are observable only when the other ionizing
processes are weak or absent, being the main ingredients of the
diffuse-ionized gas (DIG) in RGs. On the contrary, in star-forming
galaxies (SFGs) an additional important contribution to the DIG is
produced by the photons leaked from \HII\ regions
\citep[e.g.][]{sanchez21,belfiore22,lugo24}.

Understanding the ionizing sources responsible for the excitation of
the interstellar medium (ISM) in galaxies is critical not only to
characterize the ISM itself but also for  estimate accurately key
evolutionary tracers, such as the star formation rate (SFR) and
chemical abundance \citep[for recent reviews, see][]{kewley19,
sanchez21}. Optical spectroscopy, particularly
diagnostic line ratio diagrams, has long served as the main tool
to identify the dominant ionization mechanisms across galaxy
populations. Among these, the Baldwin, Phillips, \& Terlevich (BPT)
diagram \citep{baldwin81}, which compares \oiii/\Hb\ 
and \nii/\Ha\ ratios, remains the most widely used. It is assumed
that it effectively separates ionization by young massive OB stars in
\HII\ regions from harder ionization sources, such as active
galactic nuclei (AGNs) and shocks, based on their differing
line-ratio signatures \citep{osterbrock89, veil01}.

Despite its utility, the classical BPT diagram—and others like
it—faces significant limitations. First, the so-called “intermediate”
or “composite” region between star-forming and AGN-dominated zones,
defined by the demarcation lines of \citet{kauff03} and
\citet{kewley01} (hereafter K03 and K01, respectively), can be populated by systems like low-luminosity and/or metal-poor AGNs, supernova remnants, and even pure
star-forming regions
\citep[e.g.,][]{cid21, agos19, nata23}. Second, evolved ionizing
sources such as post-AGB stars or HOLMES can mimic AGN-like line
ratios—even though their emission is intrinsically weaker. Shocks,
both high- and low-velocity, can also reproduce AGN-like signatures for particular gas properties, velocity, and magnetic field strength
\citep[e.g.,][]{dopita96, carlos20}.

These poorly defined areas challenge the interpretation of diagnostic
diagrams. In response, hybrid approaches have been introduced, such as
the WHaN diagram \citep{cid-fernandes10}, which uses also the
\nii/H$\alpha$ ratio and the equivalent width of \Ha\ (\EWHa). Similar
strategies have been proposed to clean star-forming sequences from RG
contamination by incorporating \EWHa\ into traditional diagrams
\citep{lacerda18,sanchez14, sanchez18}. However, even these
improvements have limitations when emission lines are weak (e.g., in
RGs), heavily dust-attenuated, or when the S/N is insufficient for all
four lines required in BPT-style diagnostics. { To overcome these
issues some authors have explored how the combination of the \oiii/\Hb\ line ratio with other suitable spectral features (such as D4000, $g-r$ color, or the EW(H$\beta$)) can effectively discriminate between different ionizing sources \citep[e.g.][]{teim18,muno25}. Finally, \citet{whad} introduced an even more simple method, the WHaD diagram, that  combines two parameters —\EWHa\ and $\sigma_{\rm H\alpha}$— both derivable from a
single emission line. These methods significantly simplify the
classification of ionizing sources.}

All diagnostic diagrams are in practice validated using observational data (e.g., K03)
or theoretical models (e.g., K01). In essence, the distribution of known/assumed line parameters, 
of galaxies and/or regions within them whose ionization is clearly known, defines a region within the
diagram. The classification procedure is validated depending on 
how clearly the regions associated with different physical
processes are separated. This
approach was followed by most studies proposing a new diagnostic
diagram and a boundary or demarcation line to separate 
different ionizing sources
\citep[e.g.][]{baldwin81,veilleux87,osterbrock89,kauff03,kewley01,cid11,whad}.

There is a fundamental limitation in this method, as it relies on
precise knowledge of the ionization mechanism responsible for the
observed or modeled properties. Consequently, when validating against
observational data, one must assume a specific ionizing
source—typically OB stars linked to recent star formation. On the
other hand, theoretical models demand the assumption of complete
understanding of the underlying physical processes, the nature of the
ionizing sources, and the characteristics of the ionized
gas. Furthermore, using data, it is assumed that only one mechanism is
present which is intrinsically impossible in complex systems such as
galaxies \citep[e.g.][]{ARAA,sanchez21}. Finally, there are
degenerancies between different mechanisms that could produce the same
observational properties (e.g., line ratios). The ionization strength
vs. metallicity degenerancy is one of the best known ones
\citep[e.g. K01,][]{sanchez15}, but there are many others that are not
frequently considered (for instance the post-AGB/shocks/AGN
degenerancy described before). The use of spatial resolved information
that allows to explore the morphology of the ionized gas, its
kinematics and even the properties of the underlying continuum
(stellar or not) is a much better method to provide an optimal
classification of the ionizing mechanism
\citep{ARAA,sanchez21}. However, this is not possible when single
aperture spectroscopy is analyzed, like in most large galaxy surveys
\citep[e.g. SDSS, DESI][]{york00,DESI}.

The aim of this study is to perform a critical exploration of how we
interpret some of the most frequently used
diagnostic diagrams (and some others recently introduced) of galaxies in the nearby Universe extracted
from the NSA catalog \citep{blanton11}. Sub-samples are selected
to be archetypal of SFGs, RGs and galaxies hosting an AGN. We
acknowledge that the ionization mechanisms present on each of those
galaxy types may not be sharply defined (as indicated before). However, this is
indeed part of the problem to be explored, as these diagrams are
frequently used to separate between those groups without taking into
account the real mixed nature of the ionization that produces the
observed properties.

This article is organised as follows: Sec.~\ref{sec:data} presents the datasets and galaxy samples employed in this study, including the different AGN selections and additional parameters used. The analysis of the data is
presented in Sec.\ref{sec:ana}. It includes the qualitative description
of the distributions of late-type and early-type galaxies (Sec. \ref{sec:diag}) along the diagnostic diagrams explored, in contrast with that of AGN hosts (Sec. \ref{sec:diag_agn}). A quantitatively study of
how the archetypal sub-samples are classified using different
schemes is included in Sec. \ref{sec:frac}. Finally how the full sample would be
classified when using those very same schemes, including our newly proposed one
 is presented in Sec. \ref{sec:final}. In Sec. \ref{sec:dis} we discuss the results,
including a revision of the methods adopted to select the AGNs in this
study in the light of our own results (Sec. \ref{sec:bona-fide}), and
a sanity-check of how we could reproduce some well established results
when adopting our proposed classification scheme (Sec. \ref{sec:hosts}).
Finally, we present the conclusions of this study in Sec. \ref{sec:con}.

The standard $\Lambda$ Cold Dark Matter cosmology with  parameters: H$_0$=71 km s$^{-1}$ Mpc$^{-1}$, $\Omega_M$=0.27, $\Omega_\Lambda$=0.73, is assumed throughout this study, in concordance with \citep{sanchez22}.


\section{Data}
\label{sec:data}

\subsection{Galaxy sample and spectroscopic data}

We extracted our sample of galaxies from version \texttt{v1\_0\_1} of
the NSA dataset
\footnote{\url{https://www.sdss4.org/dr17/manga/manga-target-selection/nsa/}}
\citep{wake17}, which is a catalog of parameters of $\sim$600,000 nearby
galaxies ($z<$0.3) selected from the Sloan Digital Sky
Survey \citep[SDSS][]{york00}. It includes improved photometric measurements
in the SDSS $ugriz$ bands, as well as far- and
near-ultraviolet photometry ($FUV$ and $NUV$ respectively) provided by
the \textit{Galaxy Evolution Explorer} \citep[GALEX][]{mart03}, and additional parameters such as the redshift of the target,
structural and morphological information, and additional quantities
such as the stellar masses, new derivation of the Sersic indices, and
new aperture corrections applied to all photometric values accounting for the PSF differences between filters. 

Additional spectroscopic information is obtained for each galaxy in the NSA
by looking for the corresponding target listed in
the catalog of galaxy properties for SDSS-DR8 \citep{SDSS8} derived
using the MPA-JHU
analysis\footnote{\url{hhttps://www.sdss4.org/dr17/spectro/galaxy_mpajhu/}}. Among the
extracted information, the most relevant for the current
exploration are the
flux, equivalent width, and velocity dispersion from the \oiii, \nii,
\sii, \oi, \Ha, and \Hb\ emission lines. Although there are more recent
analyses of the same dataset, this one has been broadly used in 
relevant explorations such as the uncovering of the mass-metallicity
relation \citep{tremonti04}, the star-formation main sequence
\citep{brin04}, as well as the seminal exploration of the distribution of
galaxies in the BPT diagrams by K03. Finally, we
apply a cut in the redshift excluding the galaxies in the local volume (z$>$0.005) and maximizing the completeness of the NSA catalog (z$<$0.1). The cross-matched
catalog (NSA-MPA-JHU, or NMJ sample hereafter) comprises a total number of 545,548
galaxies, including all galaxy types and covering a wide range of
stellar masses. This sample could be considered by all means
representative of the population in the local Universe (once excluded dwarf galaxies), and it is one
of the largest samples of galaxies with spectroscopic information
available to date within the considered redshift range.

\subsection{AGN samples}

The selection of  bona-fide AGNs to validate
the classification using different diagnostic diagrams is a
difficult task. Thus, different approaches have been adopted in the literature. For instance, \citet{whad}
used two samples of X-ray selected AGNs from
\citet{nata23} and \citet{agos19}, under the assumption that the X-ray
emission is a reliable tracer of the nuclear activity. However, this
may bias the results towards a particular type of 
objects, so, we prefer to follow \citet{comer20} and select a
sample of AGNs based on different selection criteria, including X-ray,
infrared, UV-optical and radio selections.

\subsubsection{X-ray selected AGNs}
\label{xAGNs}

The sample of X-ray selected AGNs (X-AGNs) was extracted from the 4XMM-DR14s
catalog\footnote{\url{https://xmmssc.aip.de/cms/catalogues/4xmm-dr14s/}}
\citep{traul20}.  This is a comprehensive compilation of serendipitous
X-ray sources detected by the \textit{XMM-Newton} observatory and it covers a wide area on the sky. The catalogue includes 427,524 
sources, of which 329,972 have been observed multiple times. In total,
it lists over 1.8 million individual flux measurements across the
standard \textit{XMM-Newton} energy bands (0.2--12.0~keV).  For each
source, parameters such as flux, hardness ratio, and variability
indicators are provided. The size, spatial coverage, and unbiased selection criteria make this sample suitable for the current
exploration.

We cross-matched the 4XMM-DR14s catalog with our NMJ sample of galaxies, looking for coordinates matching within
3$\arcsec$ (i.e., the size of the SDSS fiber). We found a total of
1390 coincidences, from which we assign the X-ray properties in the
catalog to the corresponding NSA galaxies. For each galaxy
we derived (i) the X-ray luminosity (L$_X$) in the hard band (2-12 keV),
using the redshift included in the catalog, and (ii) the hardness ratio ($HR$), defined as:

\begin{equation}
HR = \frac{H-S}{H+S}
\end{equation}

\noindent
where $H$ corresponds to the flux in the X-ray hard band 
and $S$ corresponds to the soft band (0.2-2 kev).

The X-AGNs candidates were those with
L$_X > $10$^{41}$ erg s$^{-1}$ and $HR > -0.2$. It was known that a
cut in luminosity of 10$^{42}$ erg s$^{-1}$ minimizes the
contamination from any ionizing source different than AGN
\citep[e.g.][]{brig11}. However, it could exclude a considerable
fraction of AGNs too \citep[e.g.][]{nata23}. Lowering the luminosity
limit by an order of magnitude would increase the possible
contamination of other ionizing sources (e.g., SF) by just a
3$\%$. Finally, we impose a cut in $HR$ to select only the X-AGNs with
the hardest radiation \citep[e.g.][]{menl13}, what is particularly
effective to select obscured targets. Adopting these criteria we
end-up with 627 X-AGNs.

\subsubsection{IR selected AGNs}
\label{iAGNs}

The all-sky imaging survey performed by the { Wide-field Infrared Survey Explorer} \citep[WISE][]{wrig10} in four bands centered at 3.4~$\mu$m,
4.6~$\mu$m, 12~$\mu$m, and 22~$\mu$m (hereafter referred to as $W1$, $W2$, $W3$, and
$W4$) was used to define a sample of infrared selected AGNs
(I-AGNs). We perform a positional crossmatch between the NMJ catalog and
the WISE catalog using a matching radius of $6\arcsec$, the typical PSF FWHM (for the shorter WISE wavelength
bands). Almost all targets in our NMJ catalog match with a
WISE target (i.e., 541,478 matched sources). For
these objects, we adopt the profile-fit magnitudes provided by the
AllWISE catalog. There are multiple { WISE}-based color selection
methods to select AGNs
\citep[e.g.][]{wrig10,jarr11,donl12,asse18,comer20}. We followed
\citet{asse18} and \citet{comer20}, as they perform an exploration
somehow similar to the one attempted here. We adopted a criterion based on two
different color cuts for two different IR brightness ranges:
\begin{equation}
\begin{aligned}
W2>13.07 : & \, W1-W2 > 0.486 \exp \left(0.092 (W2-13.07)^2\right) \\
W2<13.07 : & \, W1-W2 > 0.486 \\
\end{aligned}
\end{equation}
This method yields a sample of 7871 I-AGNs.

\subsubsection{UV/Optical photometry selected AGNs}
\label{oAGNs}

We use the ultraviolet and optical photometry included in the NSA
catalog and select candidates to AGNs based on the color distributions
shown by \citet{tram07}, combining the following criteria: (i)
$NUV-u<$2 mag, (ii) $NUV-g<$4 mag, (iii) $FUV-NUV<$0.5 mag, (iv)
$u-g<$0.6 mag and (v) $g-r<$0.6 mag. An additional cut has been
included in the absolute magnitude of the UV bands to exclude
intrinsically faint targets ($FUV_{abs}<-$19.5 mag and $NUV_{mag}<-$18
mag). We should note that some of these criteria are somehow
redundant, and the most restrictive ones are those including the
$NUV-u$ and $u-g$ colors. A total of just 330 objects are selected
using this rather restrictive criterion, tracing essentially
unobscured AGNs (O-AGNs hereafter). The low number of recovered
O-AGNs is due to the selection criteria, that was tuned to select
QSOs, i.e., AGNs in which the contribution of the host galaxy
is negligible (on the contrary of those objects included in our NMJ sample).

\subsubsection{Radio selected AGNs}
\label{rAGNs}

We use the Faint Images of the Radio Sky at Twenty centimeters radio
survey \citep[FIRST][]{beck95} to identify a sample of radio AGNs
(R-AGNs) in the NMJ catalog. FIRST has observed 10,000 square
degrees of both hemispheres at 1.4 GHz, generating a catalog of $~$1
million sources. Following the same procedure described in
Sec. \ref{iAGNs} we cross-matched their position in the sky with that
of our main galaxy sample, looking for coincidences within the same
distance as the one adopted for selection of I-AGNs. We found a good matching for 18851
galaxies. From them we select the brightest and clearly resolved
targets by imposing the following criteria: (i) a minimum integrated
flux of 10 mJy, (ii) a minimum FWHM along the major axis of
0.5$\arcsec$, and (iii) that the integrated flux is
at least 1.1 larger than the peak flux \citep[following ][]{ivez02}. These criteria minimize the possible contamination by
star-forming galaxies that may present emission in the radio continuum, being usually fainter and more compact \citep[e.g.][]{wada04}. Similar
criteria have been adopted in the literature to select extended
radio sources \citep[e.g.][]{kimb11}. Following this procedure
we select a final sample of 1098 R-AGNs.

\begin{table}
  \caption{Number of galaxies and AGNs in the analyzed sample}
\label{tab:sample}      
\centering                          
\begin{tabular}{rrrrrr}
\hline\hline                 
         & NMJ     &  XMM  & WISE  & UV/Opt.  & FIRST \\
\hline               
\# gal.  &  545548  &  1390      & 541478  &   547928  &  18851   \\
\# AGNs  &   9449   &   627      &   7871  &      330  &   1098   \\
\% AGNs  &  1.71  &   45.11  & 1.45  & $<$0.01 &   5.82 \\
\hline                                   
\end{tabular}
\end{table}

{ 

\subsection{Final galaxy and AGN sample}

In summary, we have compiled a catalog of more than half a millon
galaxies, comprising photometric, structural and spectroscopic
properties, together with positions on the sky, redshift and
distance, by combining the NSA and MPA-JHU catalogs, what we call the NMJ
sample. In addition, we have created four samples of AGNs by (1) cross-matching NMJ with the 4XMM-DR14s catalog of X-ray
sources, applying a cut in luminosity and hardness ratio (X-AGNs,
Sec. \ref{xAGNs}), (2) cross-matching NMJ with the AllWISE catalog of infrared sources,
applying a cut in the infrared colors depending on the brightness of
the targets (I-AGNs, Sec. \ref{iAGNs}), (3) by selecting objects with a clear UV/blue color excess (O-AGNs, Sec. \ref{oAGNs}) and (4) by cross-matching the NMJ sample with the FIRST catalog of radio sources, applying an absolute and relative threshold in their extended fluxes and their projected size (R-AGNs, Sec. \ref{rAGNs}). Table \ref{tab:sample} lists the number of
objects included in the NMJ catalog and the result of cross-matching it with
each of the multiwavelength catalogs described before (XMM, WISE, UV/Opt. and FIRST), together with the number and fraction of AGNs selected using those datasets. The total
number and fraction of AGNs selected combining the four methods has
been listed too.
}

   \begin{figure}
   \centering
   \minipage{0.99\textwidth}
   \includegraphics[width=8.5cm,trim={1 5 15 8},clip]{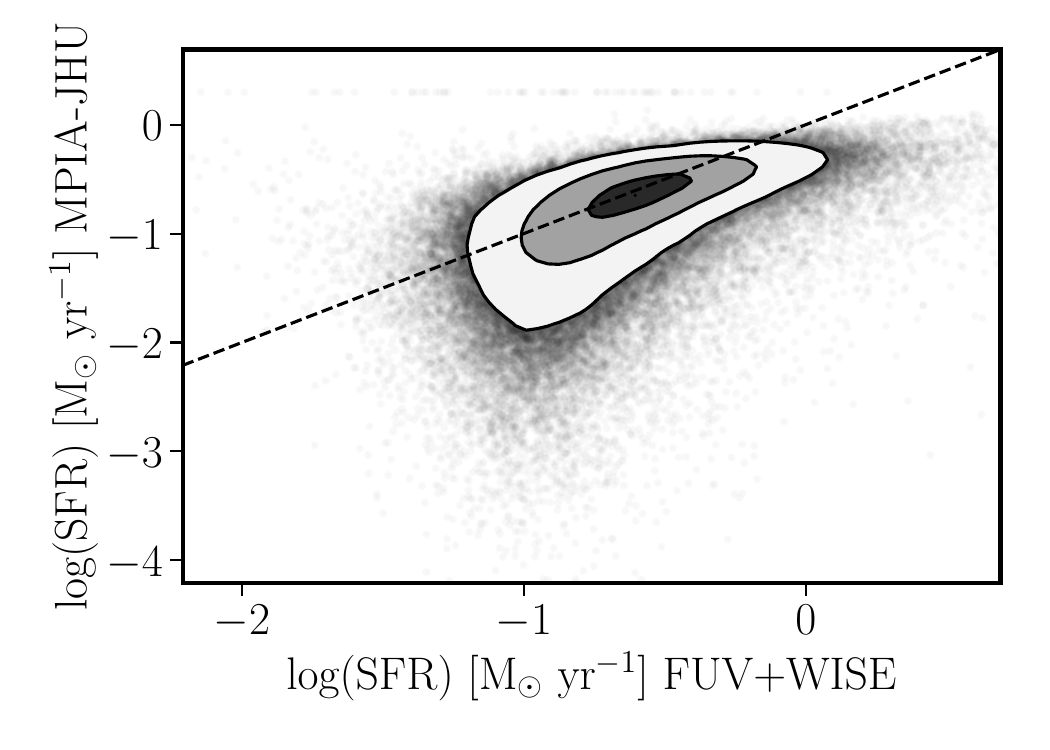}
 \endminipage
 \caption{Comparison between the SFR provided by the MPA-JHU catalog and the one derived combining the infrared and ultraviolet photometry as described in the text. The black dots correspond to each galaxy in the NMJ sample { (i.e., the full sample of galaxies analyzed in this article)} and each successive grey contour represents the area encircling a 90\%, 65\% and a 15\% of these points. The dashed-line represent the one-to-one relation. }
 \label{fig:SFR_comp}%
\end{figure}

\subsection{Additional parameters}
\label{sec:add}

The combination of the NMJ catalog with the XMM, WISE, and
FIRST datasets provides a large
sample of galaxies with a wide set of physical properties. For the purpose of this work we derive two additional parameters to characterize the explored galaxies: the disk fraction and the integrated star-formation rate.

\subsubsection{Disk Fraction ($f_{disk}$)}

It is relevant for our exploration to know whether or not a
galaxy presents a prominent disk. We  use the fact that disk-dominated/late-type galaxies and bulge-dominated/early-type galaxies are located in different regions of the effective radius (R$_e$) versus stellar-mass M$_\star$ plane \citep[e.g.][]{shen03,vand14,lang15}, defining the disk fraction, $f_{disk}$, as:
\begin{equation}
  \begin{aligned}
    R_e \ge R_{e,LT} : & \, f_{disk} = 1 \\
    R_{e,ET} < R_e < R_{e,LT}  : & \, f_{disk} = 1 - |R_{e,LT}-R_e|/|R_{e,LT}-R_{e,ET}|\\
    R_e \le R_{e,ET}  : & \, f_{disk} = 0 \\
\end{aligned}
\end{equation}
where R$_{e,LT}$ and R$_{e,ET}$ are the effective radius of the stellar mass predicted by
\citet{shen03} for late-type (LT) and early-type (ET) galaxies,
respectively. 
As a proxy of R$_e$ we adopted the {\tt ELPETRO\_THETA\_R} parameter in the NSA catalog, transformed to kpc using the angular distance estimated using the
standard cosmology and the redshift provided by the same catalog.
Finally, we use {\tt SERSIC\_MASS} in the NSA for M$_\star$

We stress that $f_{disk}$ should not be taken as a detailed estimation
of the real fraction of disk (or bulge) in luminosity or mass in a
galaxy. However, it provides a simple and robust method to
segregate between disk dominated and bulge dominated galaxies. Furthermore, it does not require a detailed profile fitting (e.g., Sersic index), neither does it rely on a discrete morphological classification.

\subsubsection{Integrated star-formation rate (SFR)}

The MPA-JHU catalog provides different estimations of the SFR. However, all
of them rely on the spectroscopic information, in particular the
H$\alpha$ flux, that is biased to the central regions sampled by
the SDSS fibers, and it may not be representative of the star-formation
state of the entire galaxy \citep[e.g.][]{rosa16,sanchez18}. To obtain  
an independent and robust estimation of the integrated SFR that takes
into account the dust obscuration, we  use of the UV and IR
photometry provided by the GALEX and WISE datasets.  Then, we adopt
the calibrators proposed by \citet{cluver17} and \citet{catalan15} to
estimate the SFR based on the 12$\mu$ and 22$\mu$ WISE photometry. We
average them to obtain a single estimation for the SFR using the two
IR bands (SFR$_{IR}$). Then, we adopt \citet{catalan15} calibrators to
estimate the SFR using the GALEX FUV (SFR$_{UV}$) and the final SFR
resulting from the combination of both SFR$_{IR}$ and SFR$_{UV}$. 

Figure \ref{fig:SFR_comp} shows a comparison between the SFR derived using
this method and the values reported by the MPA-JHU for this parameter ({\tt SFR\_tot\_p50} in that catalog) { for the full NMJ sample analyzed along this study}. Although there is a relatively good correlation between both estimations in the range of high values (following almost a one-to-one relation) the MPA-JHU reports a wider range of SFRs, with a clear trend
to much lower values, in particular in the range of low values. This is exactly what would be expected when extrapolating the H$\alpha$ emission in the center
of galaxies towards their entire extensions based on a limited aperture for intermediate type galaxies (early spirals) such as Sa/Sb morphologies, as indicated before. Additional differences
are expected as the time-scale of the star-formation sampled by the different indicators (H$\alpha$ vs. IR/UV) is intrinsically different \citep[e.g.][]{kennicut83}.

When required, we will use the M$_\star$ described in the previous section together with
the derived SFR to obtain the specific star-formation rate (sSFR$=$SFR/M$_\star$).

\section{Analysis}
\label{sec:ana}

   \begin{figure*}
   \centering
   \minipage{0.99\textwidth}
   \includegraphics[width=\linewidth,trim={10 20 20 20},clip]{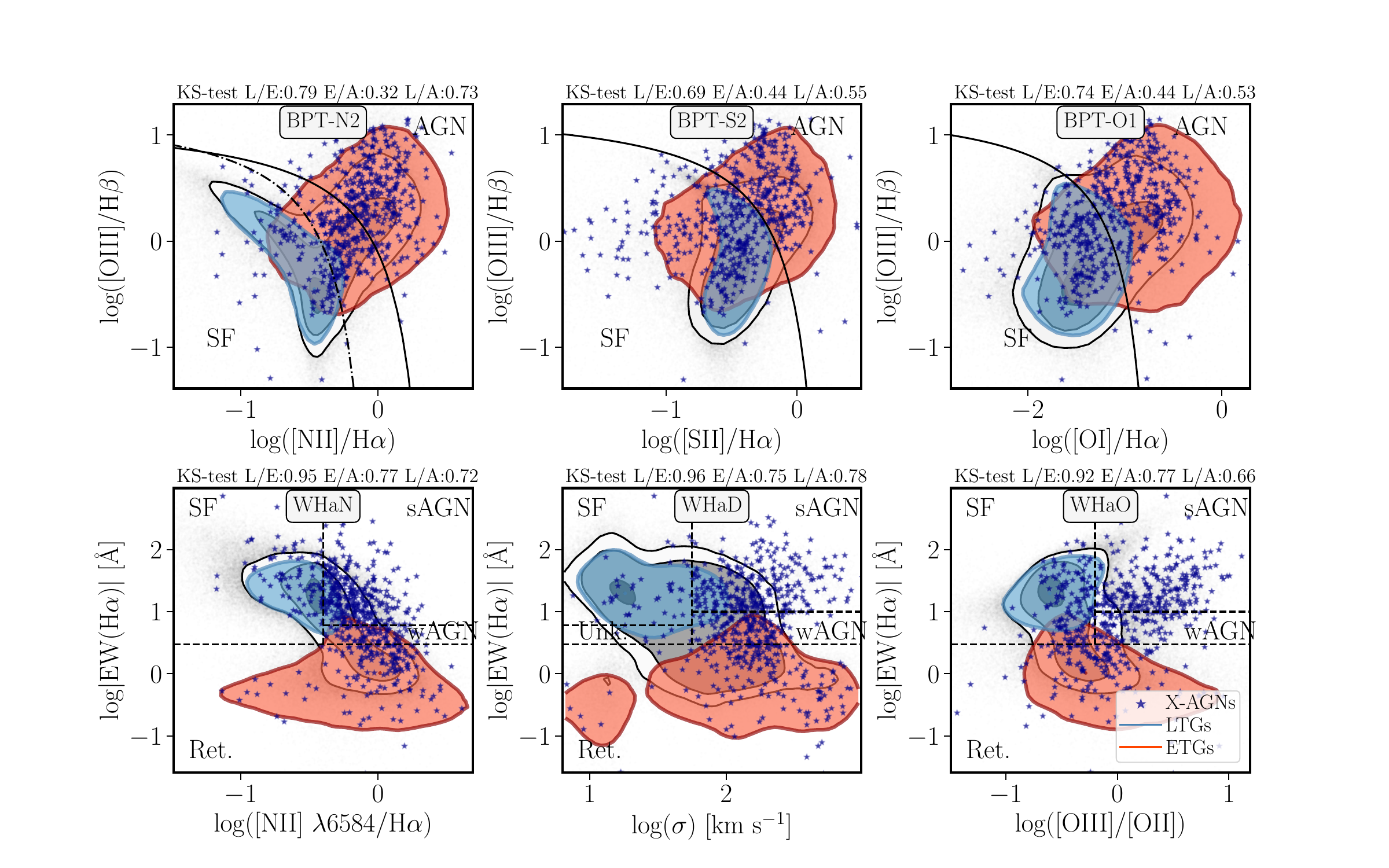}
 \endminipage
 \caption{Distribution of the sub-samples of galaxies across different diagnostic diagram. {\it Top panels:} Classical BPT diagrams \citep{baldwin81}, 
 showing the distribution of \oiii/H$\beta$ line ratio as a function of \nii/H$\alpha$ ratio ({\it left panel}), \sii/H$\alpha$ ({\it middle panel}) and \oi/H$\alpha$ ({\it right panel}). 
 Solid and dot-dashed lines correspond to the demarcation lines proposed by K03 and K01 to distinguish between the different ionizing sources. 
 {\it Bottom panels:} Diagrams comparing the distribution of Equivalent-width of H$\alpha$ (WH$\alpha$) as a function of (i) the \nii/H$\alpha$ ratio {\it left panel}, {\tt WHaN} diagram \citep[][]{cid-fernandes10},
  (ii) the H$\alpha$ velocity dispersion ($\sigma_{\rm H\alpha}$, {\it middle panel}), {\tt WHaD} diagram \citep[][]{whad}, and (iii) the \oiii/\oii line ratio ({\it right panel}), proposed here as the {\tt WHaO} diagram. In each panel the black dots correspond to the full NMJ sample and each successive grey contour represents the area encircling a 90\%, 65\% and a 15\% of these points. 
 The blue (red) contour represent the area that encircles 90\% of the values corresponding to the late-type (early-type) subsamples of galaxies, as defined in the text. 
 Finally, the location of the X-ray selected AGNs are shown as dark-blue stars. { The D-parameter derived for a set of 2D KS-tests comparing the distributions of the different subsamples are included on top of each panel, using the nomenclature L/E when comparing LTGs vs. ETGs, E/A for ETGs vs. X-AGNs and L/A for LTGs vs. X-AGNs.} }
 \label{fig:diag_XAGNs}%
\end{figure*}

As indicated in the introduction, we aim to determine whether or not the three
groups in which we divide galaxies according to their main ionization
mechanism (SFGs, RGs, and AGNs) are located in well defined regions
in a set of diagnostic diagrams. With this purpose in mind, it is
important to select three subsamples of galaxies trying to limit as much
as possible any possible contamination by the ionization dominating
the other subsamples. Furthermore, to avoid as much as possible circular arguments, 
in this selection we use galaxy
properties that are not explored by the diagnostic diagrams. 

For AGNs, we just adopt the four sub-samples described in the previous sections, as all of them
fulfill the previous requirements. As an archetypal sub-sample of SFGs, we select
blue ($u-g<$2) late-type galaxies ($n_{sersic}<$1.5), without evidence
of a bulge ($f_{disk}>$0.85), and clearly located in the
star-formation main sequence \citep[SFMS, e.g.,][]{brin04,renzini15},
i.e., log(sSFR)$>$-11.5 dex \citep[following][]{sanchez18b}. With the additional criterion that the
H$\alpha$ flux has a S/N$>$3, this sub-sample of essentially late-type
 galaxies (LTGs) consists of 90,076 objects. On the contrary, our
sub-sample of non-starforming galaxies are selected as red ($u-g>$2),
early-type galaxies ($n_{sersic}>$3.5), without clear evidence of a disk
($f_{disk}<$0.05), and well below the SFMS (sSFR$<$-11.5
dex). Considering a similar minimum S/N in the H$\alpha$ flux the
sub-sample of early-type galaxies (ETGs) comprises 43,295 objects. \footnote{We prefer to label these two samples as LTGs and ETGs, instead of SFGs and RGs, as we reserve the last terms for the galaxies selected using the diagnostic diagram.}

\subsection{LTGs and ETGs across the diagnostic diagrams}
\label{sec:diag}

Figure \ref{fig:diag_XAGNs} shows the distributions along a set of diagnostic diagrams for the entire sample of NMJ galaxies and the subsamples of LTGs, ETGs, and X-AGNs defined before, together with the boundaries defining regions associated with different physical processes. The top
panels correspond to the classical BPT diagrams
\citep{baldwin81,veil01} that present the distribution of the
\oiii/\Hb\ line ratio as a function of \nii/\Ha\ ({\tt BPT-N2}),
\sii/\Ha\ ({\tt BPT-S2}) or \oi/\Ha\ ({\tt BPT-O1}). The bottom
panels include three diagnostic diagrams that represent the equivalent
width of H$\alpha$, EW(H$\alpha$), along \nii/\Ha\
\citep[{\tt WHaN},][]{cid-fernandes10}, the H$\alpha$ velocity dispersion
\citep[{\tt WHaD},][]{whad}, and the \oiii/\oii\ line ratio ({\tt WHaO} diagram, hereafter). 
As already discussed in Sec. \ref{sec:intro}, all diagrams attempt to segregate between the ionization associated with recent SF and AGN. In addition, the diagrams using \WHa\ include a new category with the ionization associated with retired galaxies ({\tt Ret.}). Those diagrams distinguish between strong AGNs (sAGNs, \WHa$>$6\AA) and weak AGNs (wAGNs, \WHa$<$6-10\AA) too. For a more simple comparison with the classification performed using the BPT diagrams, we will not distinguish between both sub-categories of AGNs and discuss them together. Finally, there are diagrams in which a certain region is labeled as mixed/composite or with unknown ionization. For simplicity, we will consider all those galaxies together in a single category labeled as {\tt Mix/Unk}. { Each diagram shows on top the D-parameter derived from a set of 2D Kolmogorov-Smirnov (KS) tests comparing the distribution of LTGs vs. ETGs (L/E), ETGs vs. X-AGNs (E/A), and LTGs vs X-AGNs (L/A). The D-parameter from a KS-test is near zero when the two samples are derived from the same parent sample, and its is near one when derived from different parent samples. Thus, a value close to zero (one) means that the two samples are indistinguisable (clearly distinguisable). The significance of these tests is in general better than  1\%, due to the large number of objects considered in each subsample.} Similar distributions for the
other three sub-samples of AGNs described in Sec. \ref{sec:data} are
included in Appendix \ref{app:diag}.

The {\tt WHaO} is a new diagram \footnote{To our knowledge it has been used in very few occasions, and not focused on the study of different ionizing sources \citep[e.g.][]{stas15}}, that is introduced following a similar reasoning used in the {\tt WHaN} and {\tt WHaD} diagrams, comparing two parameters that trace two different physical properties associated with different ionization mechanisms: (i) the EW(H$\alpha$)
traces the relative strength of the H$\alpha$ emission line with
respect to the continuum level. High (absolute) values are found in
either galaxies under star-formation or hosting an AGN, while low
values are observed when neither star-formation nor strong AGN are
present, and (ii) the \oiii/\oii\ ratio, frequently used to estimate the
ionization parameter \citep[$U$, e.g.][]{dors11,sanchez15,espi22}, but
actually tracing better the hardness/shape of the ionizing spectrum
\citep[][]{mori16}. The harder  the ionizing spectrum (e.g., in the
case of post-AGB stars and AGNs) the larger this parameter should be.

The distributions along the different diagrams agree to the expectations and the previous knowledge
\citep[e.g.][]{ARAA}. If we focus on the {\tt BPT-N2} diagram, it is clear
that LTGs are found in the classical location of
HII regions \citep[e.g.][]{osterbrock89}, following the left-branch of
the well-known V-shaped distribution for the entire galaxy sample.
However, they present a slight shift towards the so-called mixed/intermediate region, 
between the K03 (dot-dashed) and K01 (solid) demarcation
lines. Similar distributions are seen, to some extent, in the other  
two BPT diagrams, with a stronger shift towards the
regions of higher \sii/\Ha\ and \oi/\Ha\ line ratios, slightly overpassing the K01 demarcation line (this does not happen in the {\tt BPT-N2} diagram).

On the other hand ETGs are distributed following mostly the
right-branch of the full galaxy distribution in the {\tt BPT-N2} diagram. They cover
a wide range of line ratios that expands from the location classically
assigned to strong AGNs (at the upper-right end of the diagram),
crossing the so-called intermediate/mixed regime, and expanding
clearly within the right-end of the distribution classically
associated with ionization related to SF \citep[i.e., the location of
where \HII\ regions are located, e.g., ][]{sanchez15,espi20,lugo24}.
This pattern is repeated in the other BPT diagrams ({\tt BPT-S2} and
{\tt BPT-O1}), clearly illustrating that none of those diagrams was defined
(and therefore their are not useful) to perform a segregation
between RGs and SFGs \citep[as already noticed in the literature, e.g., ][and references there in]{ARAA,sanchez21}.

\begin{figure*}
  \centering
  \minipage{0.99\textwidth}
  \includegraphics[width=\linewidth,trim={45 55 40 55},clip]{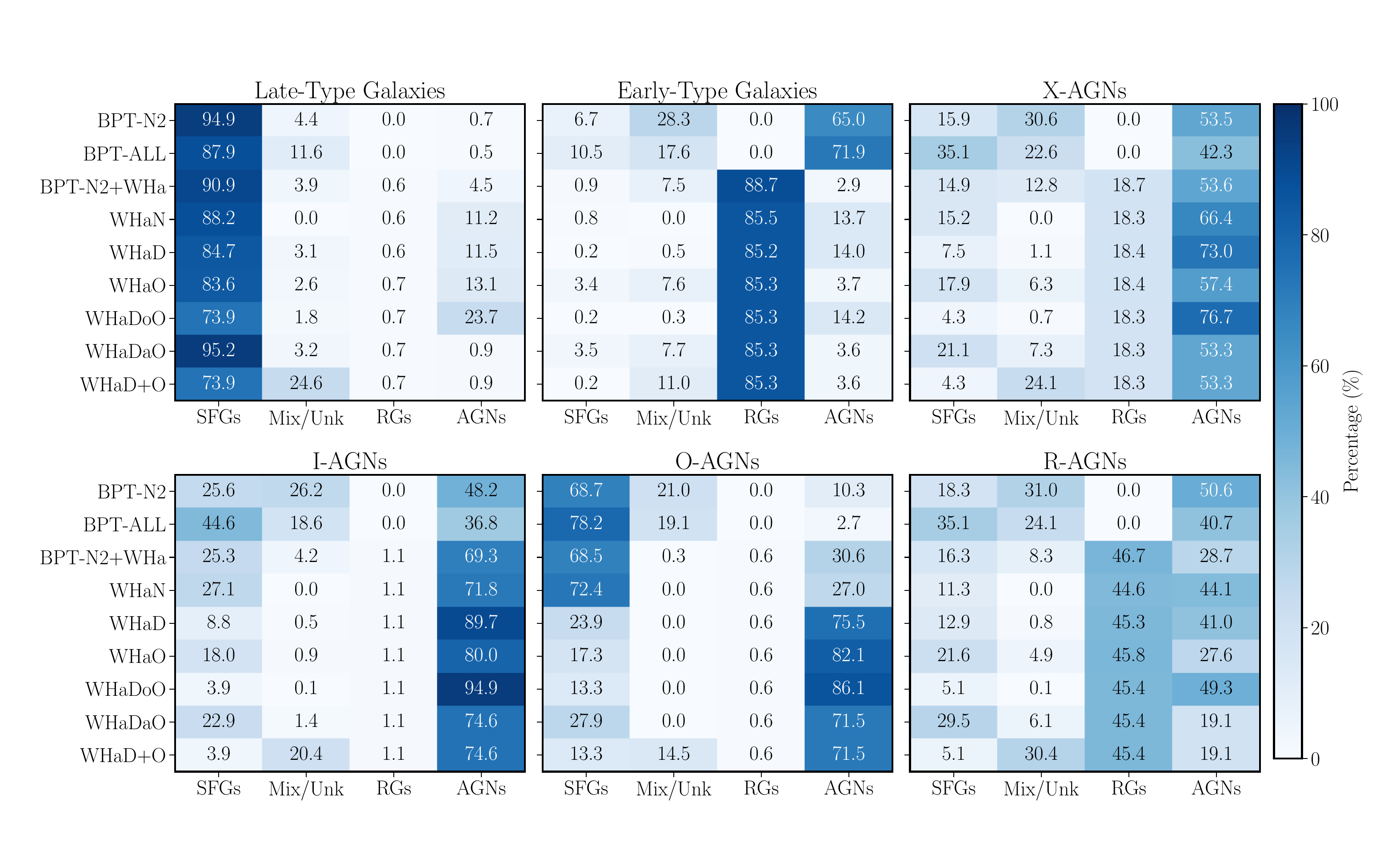}
\endminipage
\caption{Figure quantifying the differences found in the classification of the dominant ionization when using different diagnostics.
From top-left to bottom-right, each panel comprises a heat-map showing the fraction of objects (color scale and values within each cell) 
assigned to each type of ionization by a different diagnostic diagram for a different sub-sample of galaxies, including late-type galaxies, early-type galaxies, X-ray selected AGNs (X-AGNs) , 
infrared selected AGNs (I-AGNs), UV-optically selected AGNs (O-AGNs), and radio selected AGNs (R-AGNs). Each heat-map columns correspond to 
the different ionizing types considered in this work, namely (i) ionization associated with recent star formation (SFGs), (ii) mixed or unknown ionization (Mix/Unk), (iii) ionziation usually found in non-starforming/retired galaxies (RGs), 
due to hot evolved stars \citep{binette94,flor11}, and/or low-velocity shocks \citep{dopita96} and (iv) ionization associated with AGNs and or shocks 
associated with galactic scale winds \citep[e.g.,][]{carlos20} (AGNs). On the other hand, each row corresponds to a different diagnostic scheme, including the use of (i) the classical diagram
by \citet{baldwin81} that uses \oiii/\Hb\ and \nii/\Ha\ line ratios ({\tt BPT-N2}), (ii) the three diagrams by \citet{baldwin81} that use the \oiii/\Hb\ vs. \nii/\Ha\, \sii/\Ha and \oi/Ha 
line rations (BPT-all), (iii) the {\tt BPT-N2} diagram including a cut in the equivalent width of H$\alpha$ ({\tt BPT-N2+WHa}), as described in \citet{ARAA}, (iv) the {\tt WHaN} diagram that uses the 
\nii/\Ha\ and the equivalent width of H$\alpha$ ({\tt WHaN}), (v) the diagram introduced by \citet{whad} that uses \nii/\Ha\ and the velocity dispersion of H$\alpha$ ({\tt WHaD}), 
(vi) the new proposed diagram that uses \oiii/\oii and the equivalent width of H$\alpha$ ({\tt WHaO}), and three different combinations that use the {\tt WHaD} and {\tt WHaO} diagrams (vii) {\tt WHaDoO}, (viii) {\tt WHaDoO} and (ix) {\tt WHaD+O}, described in the text.}
\label{fig:heat}%
\end{figure*}

The diagrams that use \EWHa,
(lower panels of Fig. \ref{fig:diag_XAGNs}) segregate much better LTGs from RGs. 
Among the parameters used to select both sub-samples we adopted
the sSFR, that would be a
tracer of \EWHa\ if it were estimated using the SFR derived from the
\Ha\ luminosity \citep[][]{sanchez14,belfiore17a} and the
data obtained within the same aperture. For this reason we adopted a different calibrator to estimate
the SFR (Sec. \ref{sec:add}). Furthermore, we note that the separation,
although driven by \EWHa, it is somehow observed in the second
parameter adopted for each of these diagrams. Thus, on average, LTGs present lower $\sigma_{H\alpha}$, lower \nii/\Ha\ ratios and lower \oiii/\oii\ ratios than ETGs.
This is expected. First, the ionized gas in disk dominated galaxy should present a lower velocity dispersion than that observed in a bulge dominated one. Second, in the case of the line ratios, a high ionization parameter and a hard ionizing radiation field produce higher values of both line ratios \citep{stas15}, being more typical of the ionizing sources present in ETGs. However, in the case of ionizing sources associated with LTGs for low metallicities and in the presence of density-bounded \HII\ regions while the \nii/\Ha\ ratio remains low, high values of \oiii/\oii\ have been reported \citep[e.g.][]{over00,kewl13,jask13,stas15}. Thus, the first of these two line ratios seem to perform a better segregation between LTGs and ETGs than the {\tt WHaO} diagram.

{ The 2D KS-tests carried out for each diagram confirm the results outlined before. For any
  of the BPT diagrams, the D-parameter resulting from the comparison of the distributions of LTGs and ETGs is smaller   than the value found for any of the diagrams that use \EWHa. For the diagrams in the top panel of Fig. \ref{fig:diag_XAGNs}, the largest reported value is 0.79 ({\tt BPT-N2} diagram), while for the diagrams in the bottom panel, the smallest value is 0.92 ({\tt WHaO} diagram).}

\subsection{AGNs across the diagnostic diagrams}
\label{sec:diag_agn}

Once established which are the preferred location of LTGs and ETGs
in the diagrams, we explore which areas are occupied by our sample
of AGNs. Again, we start with the {\tt BPT-N2}, the most commonly used diagram to select AGNs
using optical spectroscopic data. The first obvious result from a
visual exploration is that X-AGNs are clearly not confined to the
usual region classically assigned to this kind of
ionization. Although a considerable number are located about the K01
demarcation line in this diagram, their distribution mimics that of the
ETGs, spanning through a wide range of line ratios, from high
\nii/\Ha\ and \oiii/\Hb\ values, to moderate \nii/\Ha\ and low
\oiii/\Hb\ ones. X-AGNs are found not only below the K01 curve, but
also below the most stringent K03, clearly invading the location
occupied by our sample SFGs (i.e., the classical location of \HII\
regions). This pattern is not only observed in the {\tt BPT-N2} diagram, it is also in the other two BPT diagrams, with a larger degree of
overlapping between the X-AGNs and LTGs.  \citet{nata23}
already showed a similar result for a limited sample of well
selected X-AGNs. We should note that, like in the case of this
study, the distribution of X-AGNs towards regions below the classical
demarcation lines is not correlated with the X-ray luminosity.  In
Fig. \ref{fig:diag_XAGNs} we coded the size of the figure by this
luminosity, showing that even the most luminous X-ray sources could be
located in the area below both demarcation lines (K01 and K03).

The distribution of X-AGNs in the diagrams including 
\EWHa\ provide a better separation between LTGs
and ETGs.  In the case of LTGs the separation is driven mostly by
the second parameter included in the diagram (i.e., \nii/\Ha,
$\sigma_{H\alpha}$, or \oiii/\oii).  For ETGs the
separation is driven by \EWHa\ (in {\tt WHaN} and {\tt WHaD}) or
by both parameters (in the new {\tt WHaO}
diagram). Certainly, there is no clear coincidence between the
footprints of X-AGNs and ETGs found for the three BPT diagrams.

{ As in the case of the segregation between LTGs and ETGs, presented in Sec. \ref{sec:diag},
  the 2D KS-tests comparing the distributions of those two sub-samples of galaxies with the
  distribution of X-AGNs, confirm the main results presented above. For any
  of the BPT diagrams the D-value resulting from the comparison of the distributions of ETGs and X-AGNs is considerably smaller ($\sim$0.32-0.44) than the value found for any of the diagrams that use \EWHa ($\sim$0.75-0.77). On the other hand, for LTGs, only the {\tt BPT-N2} diagram  presents a D-value (0.73) similar to the one of the diagrams using \EWHa\ ($\sim$0.66-0.78). Finally, the results from the KS-tests indicate that X-AGNs are less clearly distinguished from LTGs and ETGs tham both galaxy samples between themselves (LTGs vs. ETGs): in all diagrams the values reported for the D-parameters for the E/A and L/A cases are smaller than the ones reported for the L/E case.}

Similar results are found when exploring the distributions
of the other three AGN samples described in Sec. \ref{sec:data}
(figures are included in Appendix \ref{app:diag}), with some
significant differences: (i) in all cases, AGNs are not confined in the
region classically assigned to these objects in the BPT diagrams, with
I-AGNs and R-AGNs covering a region similar to X-AGNs,
and O-AGNs being located mostly below the K01 demarcation lines in
three BPT diagrams; (ii) I-AGNs trace better the classical loci
assigned to Seyfert-II galaxies in the BPT diagram \citep[e.g.][]{kewley06}, while R-AGNs present lower \oiii/\Ha\ values
for a given \nii/\Ha, \sii/\Ha, or \oi/\Ha\ ratio, tracing the region
usually assigned to LINERs \citep[][]{heckman87}. In both cases the distribution crosses the K01 (and K03) demarcation line invading the area associated to SF ionization; (iii) regarding the diagrams including \EWHa\, both I-AGNs and R-AGNs follow a somehow similar
pattern as the one described for the X-AGNs, with a significantly larger
number of R-AGNs located in the area covered by ETGs (in agreement
with their distribution in the BPT diagrams), and a larger number of
R-AGNs found in the area assigned to SF related ionization in the {\tt WHaO} diagram; finally, (iv) O-AGNs are located in the same region covered by X-AGNs only for the {\tt WHaD} and {\tt WHaO} diagrams, but not for the {\tt WHaN} one.
These differences reflect the different kind of AGN activity
traced when applying different selection criteria.

\subsection{Quantifying how well the ionization is classified }
\label{sec:frac}

In Sec. \ref{sec:diag} we explored the distribution of
the three samples of galaxies (LTGs, ETGs, and AGNs) in a set of diagnostic diagrams.
Here, we quantify how well the ionization can be classified based on these diagrams using
these three samples as proxies. To
do so, we adopt the following classification schemes: 

\begin{enumerate}[i)]

\item {\tt BPT-N2}: It is the most frequently adopted scheme in the literature. It uses the location across the {\tt BPT} diagram using the \nii/\Ha\ ratio to classify
galaxies as SFGs (below the K03 curve), mixed/unknown (above the K03
curve and below the K01 one), and AGNs (above the K01 curve). This scheme
is not able to select RGs by construction.

\item {\tt BPT-ALL}: Defined by the location across the three BPT diagrams, classifying the galaxies as
SFGs if they lay below the K01 curve in each diagram, AGNs if the lat above the same
curves in all diagrams and mixed/uknown if the do not fulfill any of
the two criteria. This method is more restrictive than the previous one
for AGNs, but not for SFGs. Like in case of the {\tt BPT-N2} method, the RGs type is not
considered by this classification procedure.

\item {\tt BPT-N2}+WHa): It uses the {\tt BPT-N2} diagram combined with a cut in 
\EWHa. This method was introduced by \citet{sanchez14}, and discussed
extensively in \citet{ARAA} and \citet{sanchez21}, as a method to
select RGs \citep[following][]{sta08,cid11}, while retaining
the information provided by the classical BPT diagrams. Galaxies are
classified as RGs if \EWHa$<$3\AA, irrespective of their
location within the {\tt BPT-N2} diagram. They are classified as SFGs
(AGNs) if they are located below (above) the K01 curve and 
\EWHa$>$6\AA. Galaxies not fulfilling any of the previous criteria
would be classified as mixed/unknown.

\item {\tt WHaN}: Galaxies are
classified as RGs in a similar way as the previous method (i.e., \EWHa$<$3\AA). For the
remaining classes, we consider them SFGs (AGNs) if \nii/\Ha$<$0.4
($>$0.4). This scheme is essentially the same as the one originally
proposed by \citet{cid11} for this diagram, with the only
difference that it does not separate between weak and strong AGNs.

\item {\tt WHaD}: RGs are selected in a similar way as in the
previous scheme, purely based on the value of \EWHa. However, SFGs
are separated from AGNs based on the velocity dispersion of the \Ha\
emission line, using a threshold of 57 km s$^{-1}$ as the maximum
value for star-forming galaxies.  On the contrary to the previous
method  galaxies with $\sigma_{H\alpha}$ below this
limit and intermediate \EWHa (3-6\AA) have undefined ionization
(unknown/mixed), following \citet{whad}. 

\item {\tt WHaO}: RGs are
selected in a similar way as in the two previous cases.  The SFGs (AGNs)
are selected as non-RGs that present a \oiii/\oii ratio lower (higher)
than 0.63 ($-$0.2 dex in logarithm scale). Like in the previous, case non-RGs and non-AGNs with an
intermediate value for \EWHa\ are labeled as unknown/mixed.

\item {\tt WHaDoO}: It combines the {\tt WHaD} and {\tt WHaO} diagnostic criteria. First,
RGs are selected using the same procedure described for those methods. Then 
galaxies are classified as AGNs if
they fulfill the criteria for being this type based on any of the two
schemes (i.e., {\tt WHaD} {or} {\tt WHaO}). Finally, galaxies are classified as SFGs 
if they are non-RGs, non-AGNs, and SFGs in both diagrams simultaneously.  When galaxies do
not fulfill any of the criteria they are labeled as unknown/mixed.

\item {\tt WHaDaO}: It is a variant of the previous method in which a
galaxy is classified as AGN if it fullfill this criteria using both the
{\tt WHaD} {and} {\tt WHaO} schemes. On the contrary SFGs are selected
as galaxies that are classified as this type in any of these two
schemes

\item {\tt WHaD+O}: It is a selection scheme developed based on the
results of the current analysis, in which RGs are selected in a
similar way as any of the previous schemes using \EWHa.  Finally
galaxies are labeled as SFGs (AGNs) if they are classified this way
using both the {\tt WHaD} {plus} {\tt WHaO} diagram. The remaining
galaxies are classified as unknown/mixed.
\end{enumerate}

Using these nine criteria we quantify how the galaxies on our initial archetypical subsamples (LTGs, ETGs, X-AGNs, I-AGNs, O-AGNs, and R-AGNs) are classified in the four
different categories (SFGs, Mixed/unknown, RGs, and AGNs). The result is presented in Figure \ref{fig:heat}, 
where we show the percentage of each type of galaxy classified in each ionization category according to the 
described classification schemes. For
instance LTGs, a sample of galaxies that by construction were selected
to present recent star-formation, are preferably classified as SFGs. However, there are clear
quantitative differences depending on the method. For instance, the {\tt BPT-N2} and the {\tt WHaDaO} method locate most LTGs in the SFGs group ($\sim$95\%). On
the contrary, the {\tt WHaDoO} method is the one that assigns a lower
number of LTGs to this group ($\sim$74\%). As expected no method
classifies a substantial number of LTGs as retired ($<$1\%). The number
of LTGs classified as AGNs is also low for the schemes using the BPT
diagrams ($\sim$1-4\%). The percentage increases when a single
diagram based on \EWHa\ ($\sim$11-13\%) is used, with the largest fraction
being assigned by the {\tt WHaDoO} method ($\sim$23\%). Finally both the {\tt WHaDaO} and {\tt WHaD+O} methods assign a very small fraction of LTGs to the AGN group ($\sim$1\%), providing with very similar results as the schemes using the BPT diagrams in this regards. The main difference between these two
methods is that the latter one assigns a large number of LTGs to the
unknown/mixed group ($\sim$25\%).

Larger differences are found in how each method classifies
ETGs in different ionization types. By construction, these methods that  
do not incorporate \EWHa\ do not recover RGs by construction.  
Both, {\tt BPT-N2} and {\tt BPT-ALL}, classify ETGs mostly as AGNs ($\sim$65-72\%), mixed/unknown ($\sim$18-28\%), and SFGs ($\sim$7-11\%). Those schemes that use \EWHa\ classify most ETGs as RGs ($\sim$85-88\%), with a very low number as SFGs ($<$4\%). The fraction of them classified as AGNs or without a clear classification is rather similar, ranging from $\sim$0\% to $\sim$14\%, depending on the method.

For the different sub-samples of AGNs, we find significant differences,
however we recover quantitatively the same patterns already described
in our qualitative analysis. For X-, I- and R-AGNs, those methods that
incorporate the BPT diagrams recover between $\sim$30\% (for R-AGNs)
and $\sim$69\% (for I-AGNs), with $\sim$50\% on average. The
fraction of those AGNs without a clear classification (mixed/unknown),
or even classified as SFGs, could be as large as $\sim$45\%. On the
contrary, the methods that adopt a single diagram involving \EWHa\
({\tt WHaN}, {\tt WHaD}, and {\tt WHaO}) recover larger fractions of AGNs
($\sim$60\%), with fractions as high as $\sim$90\% in some
cases (I-AGNs, {\tt WHaD}), with the sole exception of R-AGNs, in which a fraction as high as $\sim$47\% is assigned as RGs. The
O-AGNs is the group that presents the more difficulty to be classified. On the one hand, 
the fraction of them classified as AGNs is
rather low ($<$30\%) for any scheme that includes the \nii/\Ha\ line
ratio ({\tt BPT-N2}, {\tt BPT-ALL}, {\tt BPT-N2}+WHa, and {\tt WHaN}). However, for the
remaining classification methods, the fraction of recovered AGNs is
similar to the those found for the X- and I-AGNs.  Finally, for all
those methods incorporating \EWHa, appart from the {\tt WHaD+O} method
(discussed below), very few AGNs ($<$12\%) of the
different subgroups are labeled as unknown/mixed, being essentially
none in many cases.

We note that the methods combining different diagnostics
diagrams using \EWHa\ maximize the selection of particular
ionizing sources by construction: AGNs in the case of {\tt WHaDoO},
and SFGs in the case of {\tt WHaDaO}. The {\tt WHaD+O} method
described in this section is an attempt to minimize the
cross-contamination by different ionizing sources. Thus, it does not
maximize the number of neither SFGs nor AGNs, but the number of
objects for which we do not have a clear classification
(unknown/mixed). This is clearly reflected in the values shown in
Fig. \ref{fig:heat}. This could be useful in those science cases in which it
is required to exclude any possible contamination between ionizing types, obtaining incomplete
but clean categories of galaxies.

In summary, our analysis demonstrates numerically what was described qualitatively in the previous section.  First,
ionization related to recent star-formation is well identified by its
location in almost any of the explored diagnostic diagrams. However,  the ionization not related to star formation is very
differently identified by each diagram and selection scheme. On one
hand, the ionization found in RGs cannot be
identified in BPT diagrams, i.e., those not using \EWHa. On the
other hand, AGNs are more accurately traced by diagrams that combine
\EWHa with an additional observable, in particular by
the {\tt WHaD}, {\tt WHaO}, and the combination of both diagrams. Finally, BPT
diagrams erroneously assign most of the ionization found in retired
galaxies to either AGNs or unknown/mixed, with a non-negligible
pollution of the galaxies selected as SF.

\begin{figure}
   \centering
   \minipage{0.49\textwidth}
   \includegraphics[width=9cm,trim={5 5 0 0},clip]{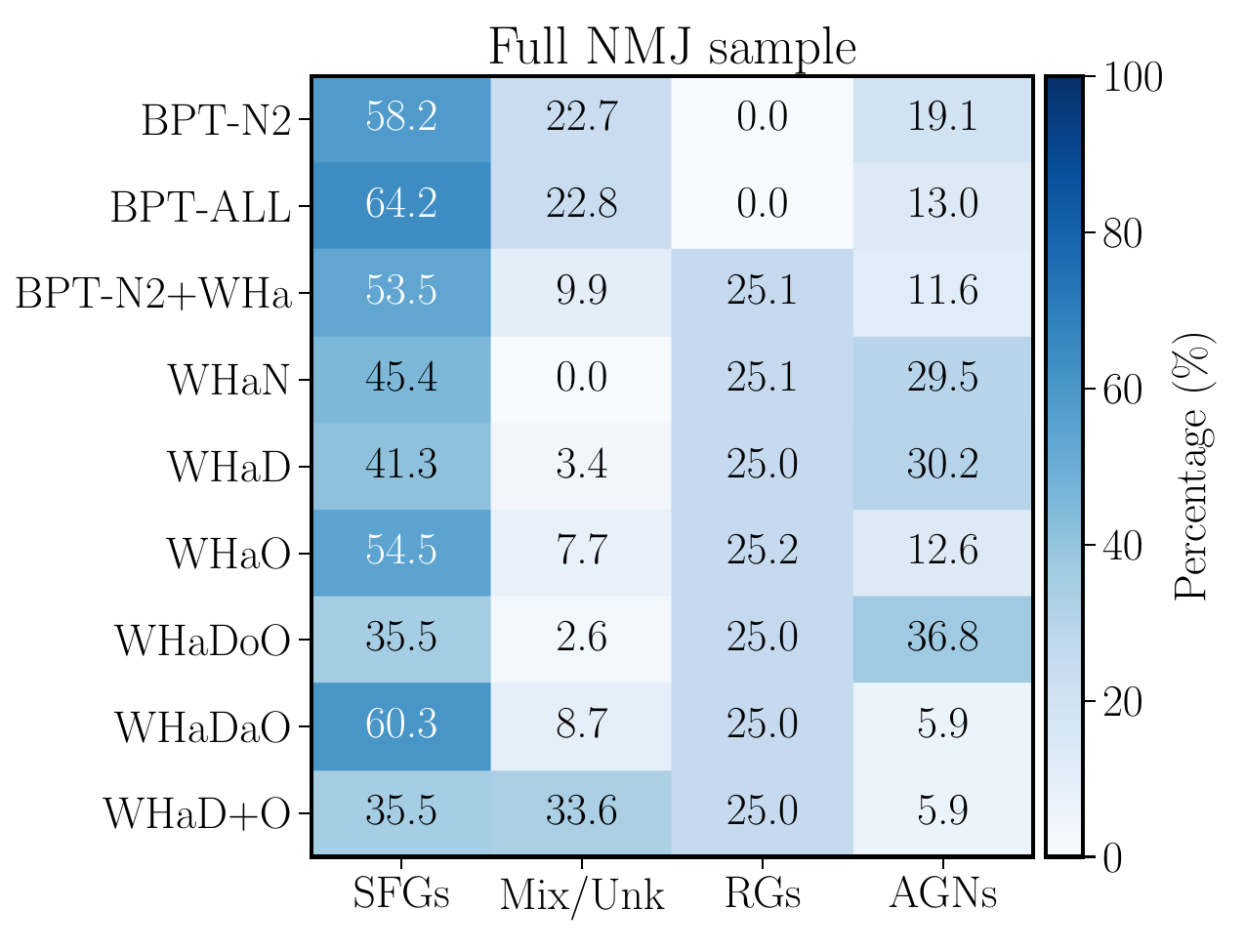}
 \endminipage
 \caption{Fraction of objects assigned to each type of ionization by 
 the different explored diagnostic schemes for the full NMJ sample analyzed along this study. Colors, labels and
 legends are the same as in Fig.\ref{fig:heat}.}
 \label{fig:heat_f}%
 \end{figure}
    %

\subsection{Classifying the ionization in the NMJ sample}
\label{sec:final}

Now, we apply the classification schemes described in the previous section to the full sample of galaxies analyzed in this study. This analysis illustrates the practical application of the
different methods. Figure \ref{fig:heat_f} shows the
distribution of galaxies along the four different ionizing groups
of this study using the classification schemes listed in
Sec. \ref{sec:frac}. Obvious differences are evident when comparing
the classification method by method. The most evident is the lack
of retired galaxies when adopting the classical BPT diagrams.  Besides
that, the fraction of both SFGs and AGNs also changes considerably.
For instance, the {\tt BPT-ALL} is the method that maximizes the number of
SFGs ($\sim$64\%), followed by the {\tt WHaDaO} (designed for this particular
purpose), while both the {\tt WHaDoO} and {\tt WHaD+O} methods
minimize the fraction of this type of ionization ($\sim$36\%). On the
other hand, {\tt WHaDoO} and {\tt WHaD} are the methods that maximize the number of AGNs ($\sim$30-37\%), while both the {\tt WHaDaO} and the {\tt WHaD+O} proposed schemes minimize them ($\sim$6\%). Finally, the fraction of RGs is essentially the same for all classification schemes that include that type ($\sim$25\%), as all of them adopt a similar approach to select them.

\begin{figure*}
   \centering
   \minipage{0.99\textwidth}
   \includegraphics[width=\linewidth,trim={5 15 15 15},clip]{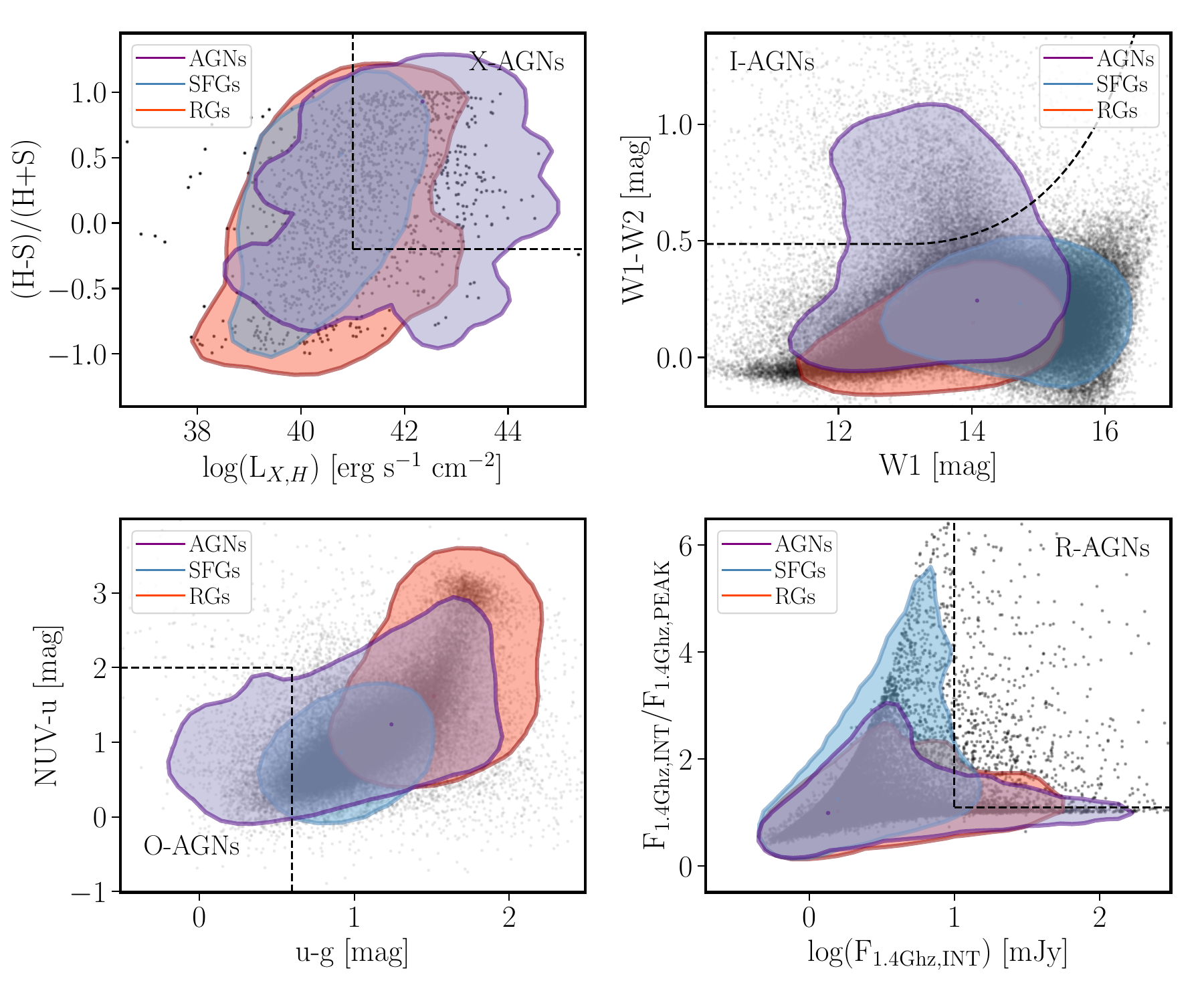}
 \endminipage
 \caption{Distribution of the full sample of galaxies in the four set of properties
 used to select the candidates to AGNs employed in this study: {\it Top-left panel:} X-ray properties, showing 
 the X-ray hardness ratio as a function of the X-ray luminosity. {\it Top-right panel:} infrared properties, showing 
 the WISE $W1-W2$ color as a function of the WISE $W2$ magnitude. {\it Bottom-left panel:} UV-optical properties, 
 showing $NUV-u$ color as a function of $u-g$ one. {\it Bottom-right panel:}
 radio properties, showing the ratio between the integrated and peak intensity at 1.4 GHz as a function of the integrated intensity. Each panel adopts the same symbols and color scheme: (i) solid circles correspond to the full sample of galaxies with measured properties, comprising 1390 objects for
 the X-ray panel, 541478 for the infrared one, 547928 for the UV-optical one, and 15839 for the radio one;
 (ii) contours represent the area that encircles 95\% of the objects with ionization classified as star-forming (SFGs, blue),
 retired galaxies (RGs, red), and AGNs (purple) using out final classification scheme described in Sect. \ref{sec:ana};
 (iii) dashed-lines show the demarcation lines described in Sect. \ref{sec:ana} to select the AGN candidates using the
 represented properties.}
\label{fig:AGN_sel}%
\end{figure*}

By comparing the different methods, we could estimate the possible
contamination between different types and possible missing
sources. {\tt WHaD} and {\tt WHaDoO} are the methods that better recover AGNs. Thus, assuming
that the fraction recovered by those methods is the closest to the real
one, then the methods based on the BPT diagrams underestimate
the fraction of AGNs by a factor between 1.5-3. The missing AGNs
are distributed in the remaining groups, contaminating them. As the
fraction of RGs recovered by all methods that include
this type is essentially the same the missing
AGNs are contaminating both the unknown/mixed group and the SFGs
type. If we adopt 10\% as the maximum fraction of unknown/mixed
ionizing sources (based on the {\tt BPT-N2}+WHa method), then it is fair to estimate that $\sim$23-25\% of the objects classified as SFGs by the {\tt BPT-N2} and {\tt BPT-ALL} diagrams most probably host an AGN. Following a similar reasoning, $\sim$3-5\% of the SFGs based on the {\tt BPT-N2} and {\tt BPT-ALL} schemes would be RGs (based on the other schemes). 
Those numbers may have an impact on the interpretation of galaxy properties (e.g., oxygen abundances) and patterns (e.g., SFMS), even though they are not particularly large (few percent), as we will discuss later.

The diagrams using \EWHa, apart from {\tt WHaDaO}, are the ones
with the lowest number of galaxies classified as mixed/unknown
($<$7\%), and a rather low number of SFGs, between $\sim$26-55\%. This
latter fraction is very similar to the one that result from the BPT
diagrams, once considering the possible contamination described
before. Thus, we conclude that they are the diagrams that provide the cleanest selection of SFGs (except the {\tt WHaO} and {\tt WHaDaO} diagrams). On the contrary, they show a non-negligible pollution in the AGN group, difficult to estimate, as they are also the diagrams that better select these targets. Being conservative, we could estimate this contamination in 
$\sim$25\%, by assuming that all galaxies classified as unknown by the
{\tt WHaO} and {\tt WHaD}oO diagrams ($\sim$8\%) are polluting the AGN group.

The {\tt WHaD+O} method was introduced to minimize the cross-contamination.
As a result, it is the method that provides the largest number
of objects with a unknown ionization ($\sim$34\%), not being 
particularly good in maximizing the recovery of AGNs ($\sim$6\%) or
SFGs ($\sim$36\%).

\begin{table*}
\caption{Number and fraction of AGNs derived using different methods, and agreement between them.}             
\label{tab:fAGNs}      
\centering                          
\begin{tabular}{rrrrrr}
\hline\hline                 
       &  NMJ    & X-AGNs & I-AGNs  & O-AGNs  & R-AGNs \\
\hline
X-AGNs & 0.11\% & 627      & 152     & 13       & 1 \\
I-AGNs & 1.44\% & 0.03\%  & 7871       & 228      & 11 \\
O-AGNs & 0.06\% & $<$0.01\%   & 0.04\%  & 330 & 228 \\
R-AGNs & 0.20\%  & $<$0.01\%   & $<$0.01\%   & 0.05\%  &  1098   \\
\hline                                   
{\tt WHaD+O}  & 5.9\%   &  53.4\%  & 74.6\% & 71.9\% & 19.1\% \\ 
\hline
\end{tabular}
\end{table*}

\section{Discussion}
\label{sec:dis}

In this study, we have explored how galaxies whose ionization is
dominated by different physical mechanisms are distributed along
frequently used and new diagnostic diagrams to evaluate how we classify the ionization using them. In particular, we selected a set of archetypal galaxies associated with recent star-formation (LTGs), the absence of
recent star-formation (ETGs), and a set of known AGNs selected using
different methods independent of the explored classification schemes. The main result of our exploration is that the most
frequently adopted procedures, based on the BPT diagrams, do not provide a robust
classification of the ionization. They maximize the number of SFGs,
polluting them with a significant number of AGNs and RG ($\sim$30\%
of the objects), neglecting the RG group, and significantly
underestimating the number of AGNs ($\sim$30\%) or missclassifying
them. A relevant result of this analysis is that there is a region in the BPT diagrams where the three archetypal groups of ionizing sources overlap: at the right-bottom end of the classical location of \HII\ regions, in the {\tt BPT-N2} diagram \citep[where more metallic regions are found][]{espi22,lugo24}. This may sound counterintuitive, as we have learned that the line ratios reflect the physical conditions of the ionized gas (e.g., metallicity, density, spatial distribution), and the properties of the ionizing source (e.g., its strength and shape). However, this relation between line ratios and physical/ionizing source properties is not univocal, and it is affected by degeneracies. We are aware of and accustomed to these degeneracies in studies of other galaxy properties, such as their stellar populations. However, they are often bypassed in the exploration of ionization.

We have several examples of very different ionizing sources that could
populate this area in the {\tt BPT-N2} diagram: (i)
high-metallicity \HII\ regions frequently found in early-spirals are
found there \citep[e.g.][]{sanchez15,espi20,lugo24}, as predicted by
well-known photoionization models \citep[e.g. K01,][]{mori16};
(ii) post-AGB ionization due to hot and low-mass evolved stars
\citep[e.g.][]{lacerda18}, also in agreement with photoionization models
 \citep[e.g.][]{mori16}; (iii) shock-ionization due to
low-scale/moderate-velocity and/or galactic-scale/high-velocity winds
\citep[e.g.][]{carlos17,carlos20}, as predicted by shock models
\citep[e.g.][]{allen08}, and (iv) AGNs, in particular bona-fide X-ray
selected ones \citep[e.g.][]{nata23}. 
We should note that AGN
photoionization models predict line ratios below both the K01 and K03
for low-metallicity AGNs \citep[e.g.][]{groves06u}. However, to our knowledge, there is no quantification of possible misclassifications and contaminations from the different sources in the literature, like the one discussed here.

Our results indicate that there is no optimal selection criterion
independent of the science case.  For instance, if the main goal of an exploration is to extract all
possible star-forming (or active galactic nuclei) irrespective of the
possible contaminations, the use of the {\tt BPT-ALL} (or the {\tt
  WHaDO}) is recommended. If, for instance, the science goal is to trace the properties of a particular population (e.g, characterizing the mass-metallicity relation or the SFMS), minimizing
the potential contamination by other selection processes, the Final scheme would be recommended. Otherwise, we may interpret as changes in the metallicity or the SFR what in reality is pollution by different ionizing processes \citep[e.g.][]{vale19}. In this sense, it is important to realize that the results and their interpretation would depend strongly on the adopted selection criteria. 
However, this is not a general conclusion either. It depends on the ionization type. We should stress that, based on our results, it is not recommended to use the classical BPT diagrams in any
exploration involving RGs and AGNs.

\subsection{How well do we select AGNs beyond diagnostic diagrams?}
\label{sec:bona-fide}

The main results from this study regarding AGNs are related to the assumptions of the methods adopted to select our archetypal subsamples: X-AGNs, I-AGNs, O-AGNs, and
R-AGNs. 
However, as in the case of diagnostic diagrams, many of the selection procedures designed using
multiwavelength photometry are based on different assumptions and the
actual knowledge and state-of-the-art at the moment when they were
developed. This has a clear impact on the number of recovered AGNs and the discrepancies in their selection of these objects using each method (as summarized in Table 2).
For instance, X-AGNs are considered the most reliable tracers of nuclear activity due to the hard X-ray emission from the hot corona around supermassive black holes, which is less affected by obscuration and orientation effects; however, X-ray surveys lack uniform sky coverage and/or completeness at faint flux levels (where the could be confused with X-ray binaries too). On the other hand, I-AGNs leverage the reprocessed emission from warm dust in the obscuring torus \citep[e.g.][]{eli06}, enabling the detection of both obscured and unobscured AGNs. Furthermore, the adopted infrared dataset has an almost uniform coverage
of the sky. 
The criteria adopted to select O-AGNs, based on the ultraviolet excess \citep[UVX,][]{sand65,schm83,boyl90}, are effective for identifying just unobscured AGNs with blue colors in color–color space \citep[e.g.][]{tram07,rich09}, but suffer from significant biases against dusty or reddened sources \citep[e.g.][]{benn98} and host galaxy contamination that may be dominant in the NMJ sample. This explains why this is the AGN sub-sample with the lowest number of objects. Lastly, radio-selected AGNs (R-AGNs) represent a distinct population characterized by synchrotron emission from relativistic jets \citep[][]{urry95}, typically found in massive elliptical galaxies and dense environments \citep[e.g.][]{sanchez99,best00}. Unlike the other groups, many R-AGNs show no optical AGN signatures and may be remnants of past activity, making them particularly challenging to classify using standard emission-line diagnostics, in particular to separate them from RGs.

It is beyond the scope of this study to revise the different procedures adopted in the literature to select AGNs considering of the current results. However, following our methodology, we explored how the different ionization types adopted in this study (SFGs, ETGs, and AGNs) are distributed in the space of parameters adopted to select the subsamples of AGNs described before.
We adopted the {\tt WHaD+O} scheme described above to segregate the 
galaxies in the NMJ catalog into the three different groups depending on the dominant ionization. This ensures the minimum cross-contamination from the different groups, at the expense of the lowest number of correctly classified galaxies.  This is a good example of a case in which pollution should be avoided, as we are to explore the typical properties of the three different types, minimizing the contamination by other types. 

Figure \ref{fig:AGN_sel}\ shows the distribution of all galaxies and the
different subsamples based on their dominant ionization in the diagrams adopted
to select X-AGNs (HR vs. L$_X$), I-AGNs (W$_1$-W$_2$ vs. W$_1$),
O-AGNs (NUV-u vs. u-g) and R-AGNs (F$_{int}$/F$_{peak}$ vs. F$_{int}$
at 1.4 GHz). A visual exploration of this figure demonstrates that there 
are significant differences among the completeness of the different methods. 
Quantitatively, $\sim$83\% of the AGNs selected by the
{\tt WHaD+O} scheme would be classified as X-AGNs (if X-ray data were
available), in contrast, only 23\% of them would be classified as
I-AGNs, and just a 1-4\% as O-AGNs and R-AGNs (if the proper data were
available). The contamination from non-AGN ionization is also
different for each selection criteria: (i) $\sim$20\% for X-AGNs
($\sim$4\% being SFGs, and $\sim$17\% being RGs); (ii) $\sim$1\% for
I-AGNs (mostly SFGs); (iii) $<$0.2\% for O-AGNs (mostly SFGs), and (iv)
$\sim$11\% for R-AGNs (mostly RGs). Thus, despite their different
ability to select complete samples of AGNs, the contamination ratio is
rather low.

By far, the most effective method seems to be the X-ray selection, although it presents the highest contamination, followed by the selection based on infrared photometry, as already shown in Tab. \ref{tab:fAGNs}. Furthermore, it presents a rather low contamination rate by non AGNs. 
On the contrary, the less effective methods are those based on UV/optical colors and radio
frequencies. These results are not surprising, as those later methods select two very particular sub-sets of AGNs: (i) unobscured AGNs in the first case, and (ii) radio-loud ones that are known
to be just a $\sim$10\% of the AGNs, when considered only the extended sources, as we did in our selection criteria \citep[e.g.,][]{urry95,raft09}.

In the light of these results, and despite the problems described
and discussed in this study, it seems that the selection of AGNs
(and other sources of ionization) using the information provided by the emission lines in the optical regime remains a powerful and efficient method compared with others proposed in the literature. This is
highlighted in Tab. \ref{tab:fAGNs}, where we include the fraction of AGNs recovered using the {\tt WHaD+O} method for the four sub-samples discussed before (X-AGNs, I-AGNs, O-AGNs and R-AGNs), and the cross-matching between them. The fraction of AGNs recovered using optical emission lines is much higher than the one recovered using any of the other four different methods using multi-wavelength observations. Indeed, there is only one AGN (candidate) that was selected using the four methods simultaneously in the entire NMJ sample. This target was also recovered using the {\tt WHaD+O} method.

\begin{figure}
   \centering
   \minipage{0.49\textwidth}
   \includegraphics[width=9cm,trim={5 35 15 0},clip]{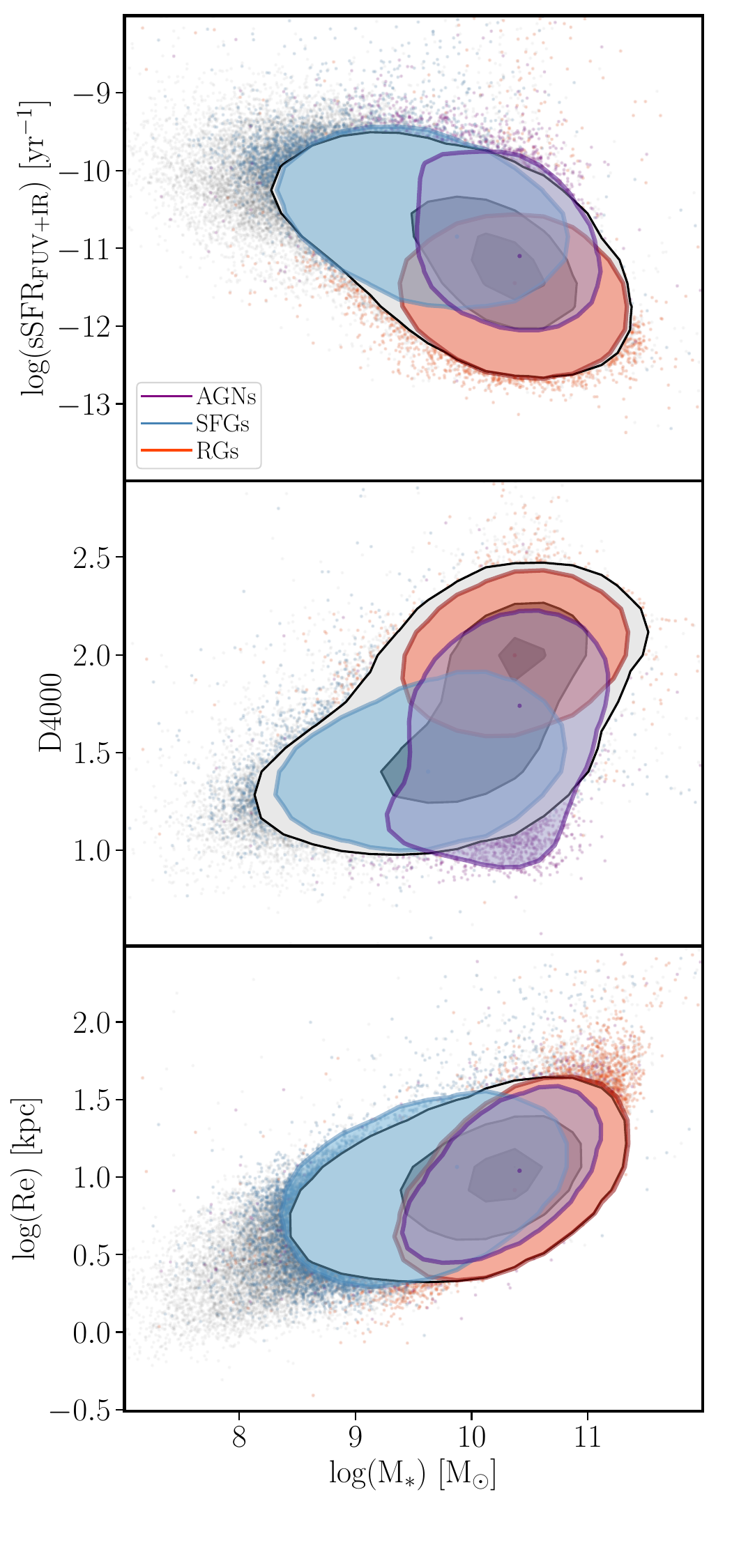}
 \endminipage
 \caption{Distribution of the full sample of galaxies along the sSFR-M$_*$ plane (top-panel), D4000-M$_*$ plane (middle-panel), and Re-M$_*$ plane (bottom-panel). Symbols and contours have the same meaning as those described in the caption of Fig. \ref{fig:AGN_sel}.}
 \label{fig:AGN_prop}%
 \end{figure}
    %

\subsection{AGN selection and the properties of host galaxies }
\label{sec:hosts}

The (proper) selection of AGNs and a good separation of them from SFGs
and RGs are relevant not only for their understanding, interesting per se, but also for studies of galaxy evolution. Nuclear activity has become extremely relevant in this
context due to the three main results:
(i) the discovery of strong correlations between black hole mass and
host galaxy properties such as bulge luminosity, mass, and velocity
dispersion \citep[see reviews by][]{Kormendy+2013,Graham2016}; (ii)
the need for an energetic mechanism—likely AGN feedback—to heat or
expel gas in massive galaxies, thus quenching star formation and reconciling the high-mass end of observed galaxy luminosity functions with theoretical predictions from semi-analytic models
\citep[e.g.,][]{Kauffmann+2000,Bower+2006,DeLucia+2007,Somerville+2008}
and cosmological simulations
\citep[e.g.,][]{Sijacki+2015,Rosas-Guevara+2016,Dubois+2016}; and
(iii) the requirement for a rapid ($\lesssim$1 Gyr) morphological
transformation from star-forming spirals to quiescent ellipticals over the last 8 Gyr, based on population studies
\citep[e.g.,][]{bell04,Faber+2007,Schiminovich+2007}.

Together, these results suggest that super-massive black holes co-evolve with
galaxies — particularly their spheroidal components
\citep[e.g.,][]{Kormendy+2013} — and that AGN feedback plays a critical
role in galaxy evolution. Specifically, negative AGN feedback may heat
or eject gas, quench star formation, and drive morphological
transitions between galaxy types
\citep{Silk+1998,Silk2005,Hopkins+2010}, potentially explaining the
evolutionary link between central and extended LI(N)ER proposed by
\citet{belfiore17a}.

Observational support for this scenario includes the finding by K03 that type-II AGNs occupy the “green valley” in the
color-magnitude diagram—between the blue cloud of star-forming
galaxies and the red sequence of quiescent ones. This has been
confirmed at intermediate redshift for type-I AGNs as well
\citep[e.g.,][]{sanchez04}, and reinforced by later studies
\citep[e.g.,][]{Schawinski+2010,torres-papaqui12,torres-papaqui13,ortega-minakata15}. AGN
hosts also appear in transitional zones of other diagrams, such as SFR
vs. stellar mass
\citep[e.g.,][]{mariana16,sanchez18,lacerda20,sanchez22}. 
Furthermore, they seem to be located in early-type massive galaxies; thus, in a
morphological transition phase between disk-dominated and
bulge-dominated galaxies.

As suggested before, all these results rely on a proper selection of
galaxies that host an AGN and a clear distinction between
galaxies that are actively star-forming or have already ceased to form
stars. It is beyond the scope of this study to explore in detail the properties of AGNs host galaxies and their connection with
galaxy evolution. However, we should at least demonstrate that our proposed {\tt WHaD+O} selection reproduces the main results described in the literature. 
To do so, we explore the distribution of our selected samples of SFGs, RGs, and AGNs extracted from the NMJ catalog using this selection criterion in three diagrams that
illustrate the evolutionary stage of galaxies: (i) the sSFR-M$_*$ diagram, which highlights whether or not a galaxy is actively star-forming 
\citep[e.g.][]{rodriguez-puebla20}, (ii) the D4000-M$_*$, illustrating the presence (or absence) of a young stellar population during a larger period than the one traced by recent star-formation
\citep[e.g.][]{blanton09}; and (iii) the R$_e$-M$_*$ diagram, that traces the compactness of a galaxy, tracing whether it is dominated by a disk or a bulge \citep[e.g.][]{hash20}.

The results of this exploration are shown in Figure
\ref{fig:AGN_prop}, illustrating that in general our proposed {\tt WHaD+O}
selection replicates previous results. In the sSFR-M$_*$ diagram the SFGs
follow a clear trend, with higher (lower) sSFR at lower (higher)
masses, expanding up to M$_*<$10$^{10.5}$M$_\odot$. 
On the contrary
RGs cover a range of M$_*$ that overlaps with SFGs above
10$^{9.5}$M$_\odot$, but covering a much lower regime of sSFRs,
distributing themselves as a cloud rather than following a clear
sequence. As expected from the literature, AGN hosts are located in the
knee/transition regime between the other two galaxy groups, following somehow the
same trend found for the SFGs, but at a mass regime covered
by the RGs.

The D4000-M$_*$ shows similar results, with SFGs showing lower D4000 values than RGs, highlighting the presence of a young stellar population that is absent in this
latter group. AGN hosts are clearly located in the transition phase
between both groups, with M$_*$ covering the highest value end of SFGs
and overlapping with those of RGs, while D4000 covers a wide range of
values, representative of both young and old stellar
populations. Thus, AGN hosts seem to be under transition between SFGs and RGs for a time larger than the usually assumed time-scale of an active nucleus \citep[e.g.][]{sanchez18,lacerda20}.

Finally, the R$_e$-M$_*$ diagram shows the clear morphological distinction
between SFGs, which are found at the expected location of
disk-dominated galaxies, and RGs, that trace the location of
bulge-dominated galaxies \citep[e.g.][]{shen03}. 
To our knowledge, this diagram has not been explored in the context of AGN hosts; however, their distribution is not surprising, as they are located again between SFGs and RGs. They follow a relation with the same slope as the one traced by bulge-dominated galaxies, but slightly
shifted towards lower M$_*$.

\section{Conclussions}
\label{sec:con}

We have evaluated how the diagnostic diagrams
classify the dominant physical processes that ionize the ISM in
galaxies. In summary, we found that:

\begin{itemize}

\item Classification schemes that rely solely on the classical BPT
  diagrams systematically \emph{over-estimate} the number of
  star-forming galaxies, \emph{under-estimate} the number of AGNs, and
  cannot recognise RGs ionisation. Quantitatively, BPT-based selections miss $\sim$30\% of
  bona-fide AGNs and mis-classify a similar fraction of RGs and AGNs
  as star-forming systems.
    
\item RGs can only be isolated robustly in the
  H$\alpha$ equivalent width; traditional BPT boundaries leave them
  hidden among AGNs or composite objects.
    
\item Diagnostics that couple EW(H$\alpha$) to an
  additional observable, e.g.\ WHaN, WHaD, and the new WHaO
  diagram, provide a far cleaner separation of ionisation mechanisms.
  In particular, the combination of WHaD and WHaO recovers
  $\gtrsim60$–90\% of independently selected AGNs while keeping SF
  contamination below $\sim$10\%.
    
\item A final, balanced selection recipe that (i) identifies RGs with
  EW(H$\alpha$)<3 Å, and (ii) labels galaxies as SF or AGN only when
  {both} WHaD and WHaO concur, yields the lowest
  cross-contamination and reproduces the expected loci of SF, RG, and
  AGN hosts in sSFR–$M_\star$, D\textsubscript{4000}–$M_\star$, and
  $R_{\rm e}$–$M_\star$ diagrams.
    
\item Multi-wavelength AGN samples (X-ray, IR, UV/optical, radio)
  occupy partly disjoint regions of optical diagnostic diagram; this
  diversity explains why any \emph{single} optical criterion alone
  cannot catch all flavours of AGN. 
    
\end{itemize}

A final conclusion of this exploration is that we should re-evaluate
carefully how we classify the ionization in galaxies, and in particular
critically revise the results presented in the
literature using the classical diagnostic diagrams and the somehow
ionization types derived from them.

{ Furthermore, following \citet{whad} and \citet{sanchez21}, we
  will attempt to implement the current methodology to
  existing IFS datasets, to explore how the use of spatially
  resolved information would improve the classification
  of the ionizing sources in galaxies.}


   \begin{acknowledgements}

    { We thanks the anonymous referee for the comments that have improved this manuscript.}
      
     SFS thanks the support by UNAM PASPA – DGAPA and the SECIHTI CBF-2025-I-236 project. Authors acknowledge financial support from the Spanish Ministry of Science and Innovation (MICINN), project PID2019-107408GB-C43 (ESTALLIDOS).
     JSA acknowledges support from the EU UNDARK project (A way of making Europe: project number 101159929)
     EP acknowledges support from the Spanish MICINN funding grant PGC2018-101931-B-I00 and Severo Ochoa grant CEX2021-001131-S funded by MCIN/AEI/10.13039/501100011033. OGM thanks the support by DGAPA-PAPIIT IN109123 and SECIHTI CF2023-G100 projects.

    We thank the creators of the NSA, MPA-JHU and SDSS surveys.
     
    The NASA-Sloan Atlas was created by Michael Blanton, with extensive help and testing from Eyal Kazin, Guangtun Zhu, Adrian Price-Whelan, John Moustakas, Demitri Muna, Renbin Yan and Benjamin Weaver. Renbin Yan provided the detailed spectroscopic measurements for each SDSS spectrum. David Schiminovich kindle provided the input GALEX images. We thank also Nikhil Padmanabhan, David Hogg, Doug Finkbeiner and David Schlegel for their work on SDSS image infrastructure.
     
    The MPA-JHU catalog was collected by a team made up of Stephane Charlot, Guinevere Kauffmann and Simon White (MPA), Tim Heckman (JHU), Christy Tremonti (University of Arizona - formerly JHU) and Jarle Brinchmann ( Centro de Astrofísica da Universidade do Porto - formerly MPA).     
     
    SDSS-III is managed by the Astrophysical Research Consortium for the Participating Institutions of the SDSS-III Collaboration including the University of Arizona, the Brazilian Participation Group, Brookhaven National Laboratory, University of Cambridge, University of Florida, the French Participation Group, the German Participation Group, the Instituto de Astrofisica de Canarias, the Michigan State/Notre Dame/JINA Participation Group, Johns Hopkins University, Lawrence Berkeley National Laboratory, Max Planck Institute for Astrophysics, New Mexico State University, New York University, Ohio State University, Pennsylvania State University, University of Portsmouth, Princeton University, the Spanish Participation Group, University of Tokyo, University of Utah, Vanderbilt University, University of Virginia, University of Washington, and Yale University.
    
    This research has made use of data obtained from the 4XMM XMM-Newton serendipitous stacked source catalogue 4XMM-DR14s compiled by the institutes of the XMM-Newton Survey Science center selected by ESA.
    
    This work is based on data from the Faint Images of the Radio Sky at Twenty Centimeters (FIRST) survey, obtained from the December 2014 data release (first\_14dec17.fits.gz), available at https://third.ucllnl.org/cgi-bin/firstcutout.
    

    \end{acknowledgements}

%
%
\bibliographystyle{aa}
\bibliography{my_bib} 


\begin{appendix}

\section{Diagnostic diagrams for different AGN selections}
\label{app:diag}

\begin{figure}
  \centering
  \minipage{0.99\textwidth}
  \includegraphics[width=\linewidth,trim={10 20 20 20},clip]{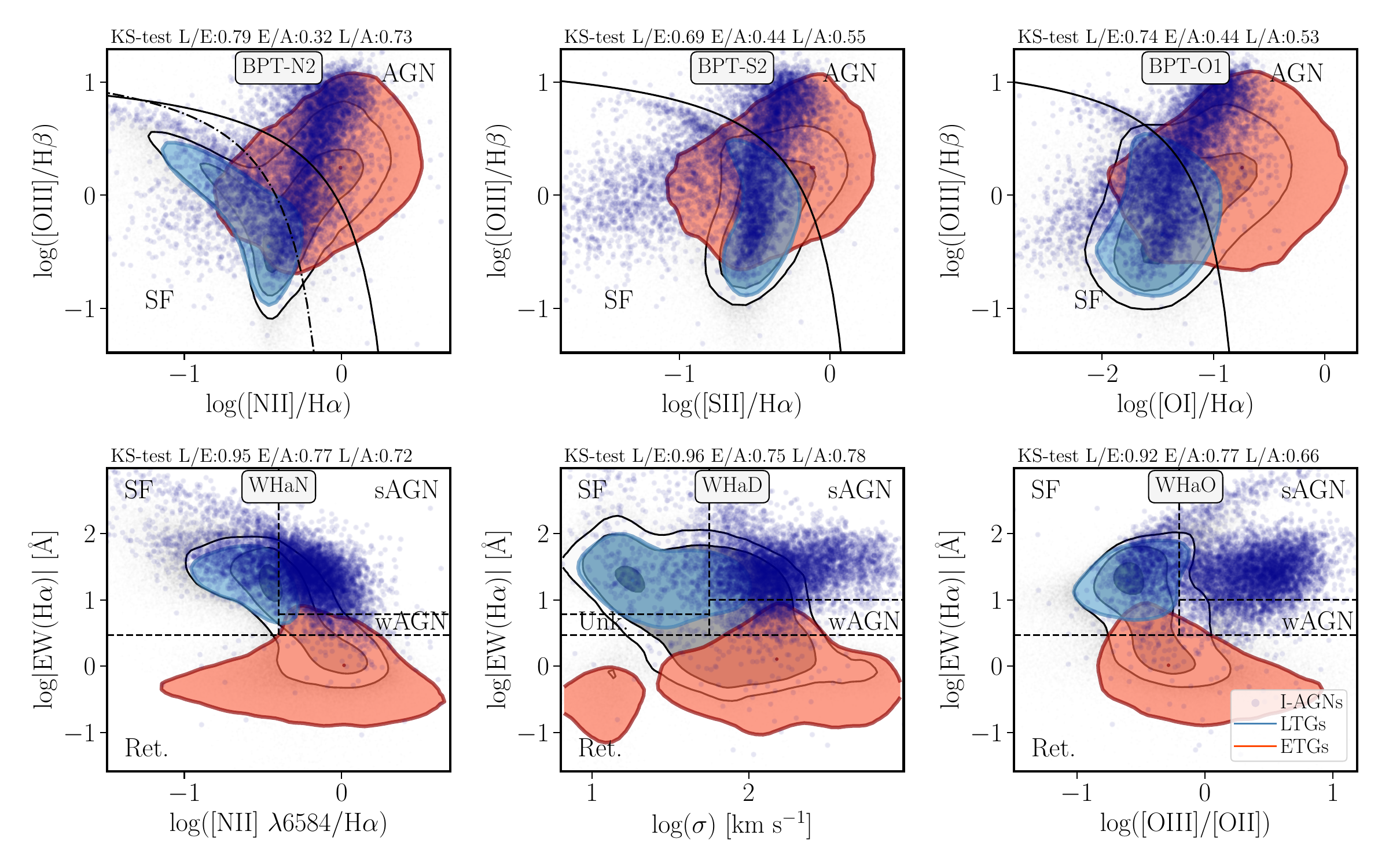}
\endminipage
\caption{Similar figure as Fig. \ref{fig:diag_XAGNs} showing the distribution of the infrared selected AGNs (I-AGNs) in the different panels as dark-blue solid circles.}
\label{fig:diag_IAGNs}%
\end{figure}

\begin{figure*}
  \centering
  \minipage{0.99\textwidth}
  \includegraphics[width=\linewidth,trim={10 20 20 20},clip]{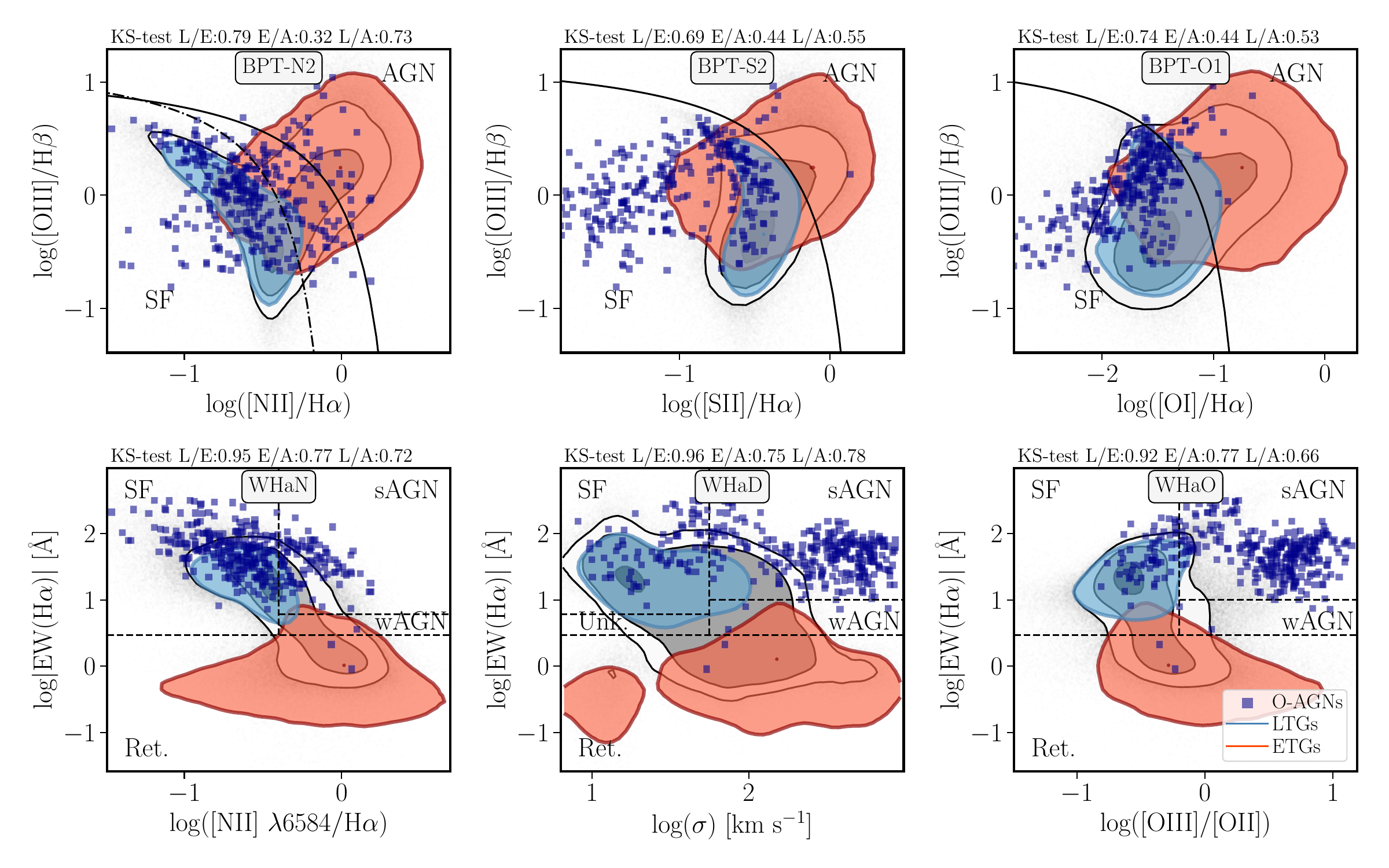}
\endminipage
\caption{Similar figure as Fig. \ref{fig:diag_XAGNs} showing the distribution of the optically selected AGNs (O-AGNs) in the different panels as dark-blue solid squares.}
\label{fig:diag_OAGNs}%
\end{figure*}

\begin{figure*}
  \centering
  \minipage{0.99\textwidth}
  \includegraphics[width=\linewidth,trim={10 20 20 20},clip]{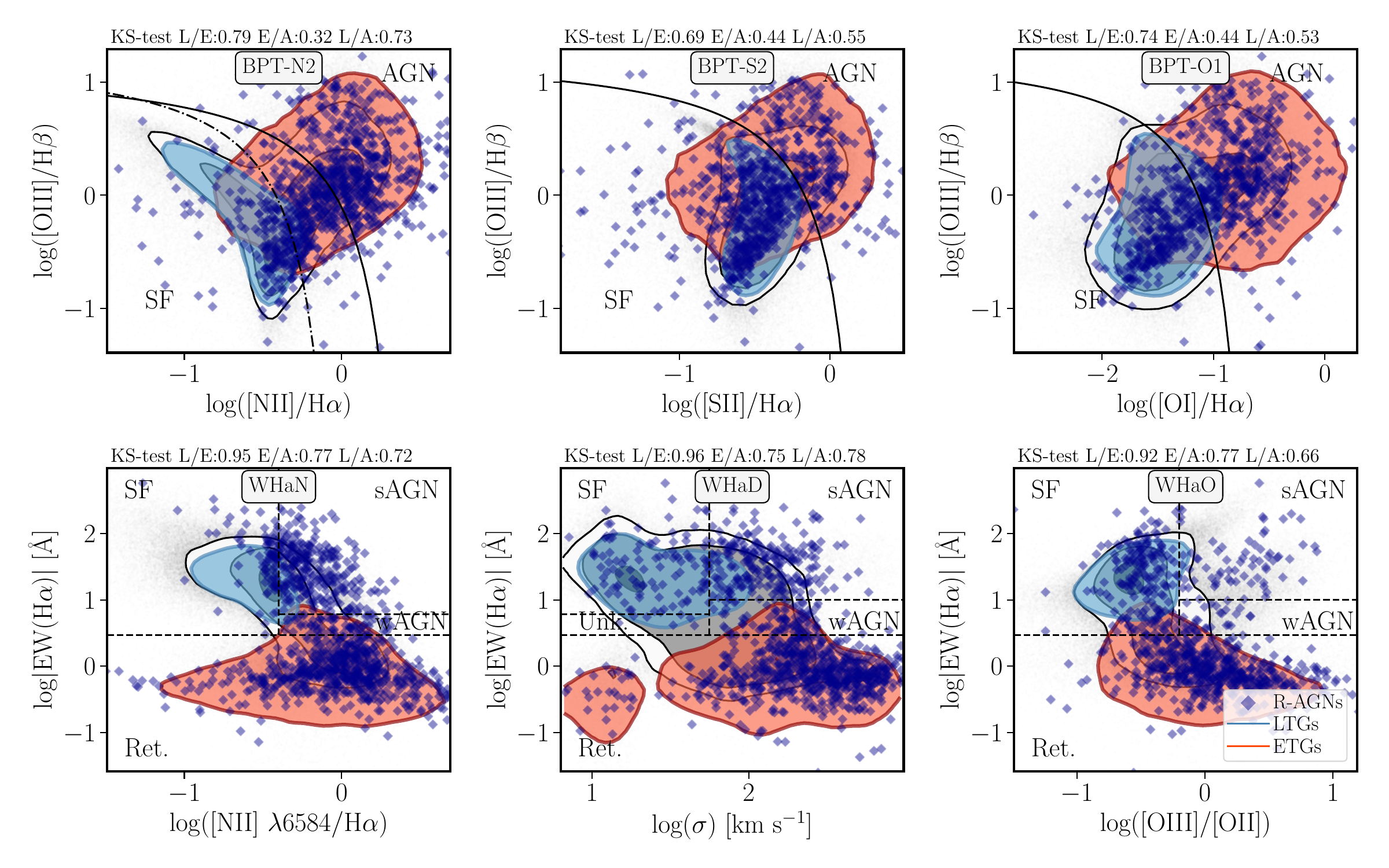}
\endminipage
\caption{Similar figure as Fig. \ref{fig:diag_XAGNs} showing the distribution of the radio selected AGNs (R-AGNs) in the different panels as dark-blue solid diamonds.}
\label{fig:diag_RAGNs}%
\end{figure*}

We present in this appendix the same diagnostic diagrams shown in
Fig. \ref{fig:diag_XAGNs} where it was compared the distribution of X-AGNs
with that of the full sample of galaxies (NMJ) and both the
two subsamples of late-type and early-type galaxies, archetypal
of star-forming and retired galaxies, corresponding to the other
three samples of AGNs explored along this study: (i) I-AGNs (Figure \ref{fig:diag_IAGNs}, (ii) O-AGNs (Figure \ref{fig:diag_OAGNs}) and (iii) R-AGNs (Figure \ref{fig:diag_RAGNs})

   %

\end{appendix}

%

\end{document}